\documentclass[twocolumn]{aastex6}
\usepackage[utf8x]{inputenc}
\usepackage{natbib}
\usepackage{graphicx}
\usepackage{multirow}
\usepackage{mathrsfs,amssymb}
\usepackage{amsmath}

\newcommand       \AU           {\,{\rm AU}}

\newcommand       \gcmthree    {\,{\rm g~cm^{-3}}}

\newcommand       \K            {\,{\rm K}}

\newcommand       \Myr          {\,{\rm Myr }}

\newcommand       \um           {\mu{\rm m}}

\newcommand       \Msun         {\,{M_\odot}}
\newcommand       \Msunyr       {\,{M_{\rm \odot}\rm yr^{-1}}}
\newcommand       \Lsun         {\,{L_\odot}}
\newcommand       \Rsun         {\,{R_\odot}}
\newcommand       \Rstr        {R_\star}
\newcommand       \Lstr        {L_\star}
\newcommand       \Tstr        {T_\star}

\begin{document}
\pagestyle{plain}
\pagenumbering{arabic}

\title{Probing Planet Forming Zones with Rare CO Isotopologues}

\author{Mo Yu \altaffilmark{1}}
\author{Karen Willacy \altaffilmark{2}}
\author{ Sarah E. Dodson-Robinson \altaffilmark{3}}
\author{ Neal J. Turner  \altaffilmark{2}}
\author{Neal J. Evans II \altaffilmark{1}}

\altaffiltext{1}{Astronomy Department, University of Texas, 1 University Station C1400, Austin, TX 78712, USA}
\altaffiltext{2}{Mail Stop 169-506, Jet Propulsion Laboratory, California Institute of Technology, 4800 Oak Grove Drive, Pasadena, CA 91109}
\altaffiltext{3}{University of Delaware, Department of Physics and Astronomy, 217 Sharp Lab, Newark, DE 19716  }

\begin{abstract}

The gas near the midplanes of planet-forming protostellar disks
remains largely unprobed by observations due to the high optical depth
 of commonly observed molecules such as CO and
H$_2$O. However, rotational emission lines from rare molecules may
have optical depths near unity in the vertical direction, so that the
lines are strong enough to be detected, yet remain transparent enough
to trace the disk midplane. Here we present a chemical model of an
evolving T-Tauri disk and predict the optical depths of rotational
transitions of $^{12}$C$^{16}$O, $^{13}$C$^{16}$O, $^{12}$C$^{17}$O
and $^{12}$C$^{18}$O. The MRI-active disk is primarily
heated by the central star due to the formation of the dead zone. CO
does not freeze out in our modeled region within $70\AU$ around a sunlike
star. However, the abundance of CO decreases because of the formation 
of complex organic molecules (COM), producing an effect that can be 
misinterpreted as the ``snow line''. These results are robust to variations in 
our assumptions about the evolution of the gas to dust ratio.
The optical depths of low-order rotational 
lines of C$^{17}$O are around unity, making it possible to 
see into the disk midplane using C$^{17}$O. Combining observations with 
modeled C$^{17}$O$/$H$_2$ ratios, like those we provide, 
can yield estimates of protoplanetary disks' gas masses.

\end{abstract}


\maketitle

\section{Introduction}

The mass and surface density of the inner $30 \AU$ of protoplanetary
disks are critical parameters that control disk evolution and planet
formation. Disk masses are currently determined by (sub)millimeter
observations of dust, because the optical depth of continuum emission is
low at radii much larger than $10\AU$
\citep{Williams_Cieza_2011}. However, observations begin to lose
sensitivity to grains as they grow beyond the observing wavelength;
millimeter wave observations may thus underestimate solid masses
in the inner disk, where dust settling and higher densities lead
to more rapid growth \citep{Perez_2012_AS209}.

On the other hand, commonly observed molecular lines often suffer from
high optical depth, and are therefore unable to probe gas near the
midplane. Molecules that are important mass contributors to outer
Solar System objects, such as H$_2$O and NH$_3$, freeze out very close
to the central star from the perspective of observers $>$ $100$ pc
away.  Thus, gas phase lines are hard to observe and most do not probe
the giant planet-forming region.  Here we seek a molecule that can be
used to probe the conditions in planet-forming midplanes with
rotational transitions observable from the ground. It must be present
in the gas phase, even in a cold disk, and be optically thin enough
that the disk is transparent in its rotational line emission.  CO is
volatile enough not to freeze out within a few tens of AU of the
central star, so is able to probe a large disk
area. \cite{AliDib_UN_formation_2014} even find that, like H$_2$O, CO
may be an important planet-building molecule for ice giants.  While
rotational emission from $^{12}$C$^{16}$O is optically thick due to
the molecule's high abundance, rare isotopologues of CO may provide
optically thin lines due to their low abundances
\citep{Miotello_2014}. If so, CO isotopologue observations could lay
the foundation for empirically determining the mass available for gas
giant formation in nearby disks.

A clue that overall disk masses may be much larger than the MMSN
(minimum-mass solar nebula) came from \citet{Bergin_HD_2013}, who
measured the mass of TW Hya using the optically thin J $=1\to 0$
transition of HD with the Herschel Space Observatory.  The $10
\Myr$-old transitional disk is estimated to have a mass of $0.05
\Msun$, much heavier than the $0.01 M_{\odot}$ of the MMSN, and larger
than many measurements estimated from dust emission. This study has
demonstrated the importance of measuring gas mass directly. However,
the HD J $=1$ state has a high excitation energy of $E/k_b = 128.5$~K
($k_b$ is the Boltzmann constant), so the line intensity depends on
not only the mass available but also the temperature of the
gas. Moreover, this transition is not accessible from the
ground. Since there is no far-infrared space mission available, this
method cannot be generally applied in the near future.

Extrapolating from line intensities to total disk mass requires
knowledge of both the abundances and the isotopic ratios in the
molecules being observed. Abundances are not constant throughout the
disk, as freezeout rate, UV intensity, and X-ray intensity all change
with distance from the star. Some chemical reactions are
isotope-selective, so the isotope ratio in molecules is also a
function of distance from the
star. \cite{Miotello_2014} have shown that the abundance ratio of 
rare isotopologues of CO to $^{12}$C$^{16}$O (If not specified, CO means 
the most common isotopologue of CO - $^{12}$C$^{16}$O  - hereafter) can deviate from the 
atomic ratio significantly due to isotope-selective photodissociation; the 
extent of the effect depends on the strength of UV radiation and dust properties.
Finally, disk temperature and density change over time, bringing
changes in the molecular inventory as the disk evolves. Here we
present chemical evolution models of protostellar disks to build a
framework for translating observed fluxes into mass densities
available for planet formation.

\begin{figure*}[ht]
  \centering
  \includegraphics[width=\textwidth]{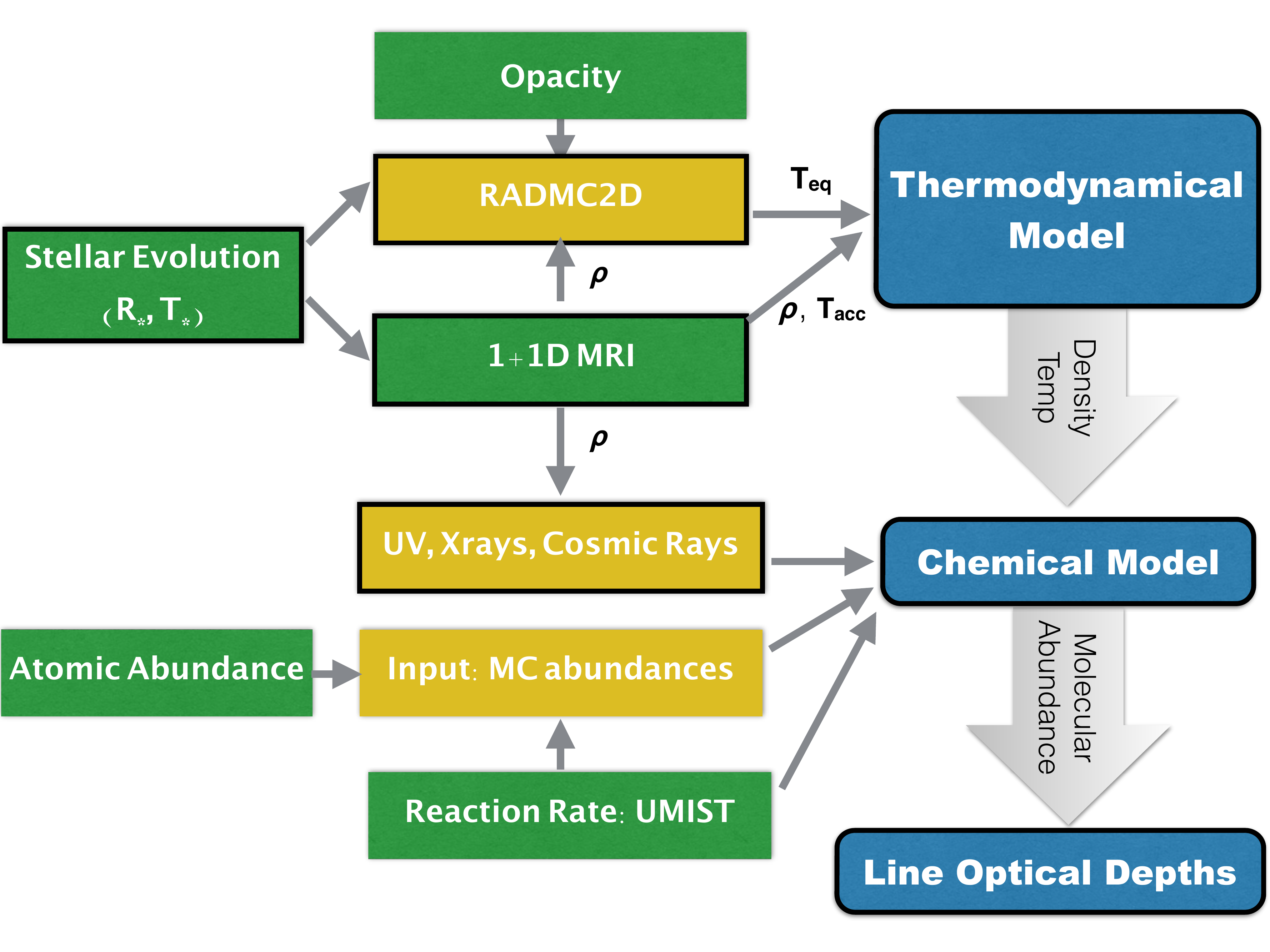}
  \caption{Workflow of our disk model\\
Blue boxes show three major components of our disk model. Green
boxes show the model inputs taken directly from the literature,
and yellow boxes are showing intermediate results computed in our
study.  Quantities in boxes with black borders evolve as a function of time, and the quantities in boxes
without borders do not change over time.\\
The arrows and variables next to the arrows are showing how
information is passed between different model components. 
\label{fig:ModelFlowChart}}
\end{figure*}

The disk model consists of three main components---the thermodynamic
model, the chemical model, and the optical depth estimation. The
workflow of the disk model is sketched in
Fig.~\ref{fig:ModelFlowChart}, and the structure of this paper follows
the model workflow.  The thermodynamic model (Section $2$) calculates
the evolution of the disk thermal structure for $3$~Myr by adding the
passive heating from the stellar irradiation to the viscous heating
from the magnetorotational instability (MRI) turbulence
\citep{Balbus_Hawley_1991}. We construct $2$D dust radiative transfer
models with RADMC \citep{Dullemond_Dominik_2004} to calculate the
passive heating, and adopt the viscous heating profile from an
MRI-active disk model \citep{Landry_2013}, which includes a viscosity
prescription for accretion driven by MRI turbulence. The chemical
model (Section $3$) then uses the density and temperature profiles
from the thermodynamic model as input to calculate the disk's chemical
evolution for $3\Myr$.  After analyzing chemical model results and
discussing the model dependencies on initial molecular cloud
abundances, uncertainties in reaction rates, and 
different grain evolution scenarios
in section $4$, we estimate optical depths of various
rotational emission lines from CO isotopologues (Section $5$), and
calculate the optical depths contributed by dust at corresponding
locations. Finally, we summarize the main results in section $6$.
A list of symbols and definitions used can be found in Table \ref{table:VarList}.

\section{The Thermodynamic Model}
\label{sec:thermodynamic} 

The thermodynamic model provides density and temperature evolution
profiles for both the chemical model and the optical depth
calculation. Our thermodynamic model is built upon the $1+1$D disk
model by \cite{Landry_2013}, who calculated the structure and
evolution of a disk with $0.015\Msun$ within $70\AU$ of the star. The
central star follows the evolutionary track of a $0.95\Msun$ star
\citep{D'Antona_Mazzitelli_1994} from $0.1\Myr$ ($\Tstr=4600K$, and
$\Rstr=5.5\Rsun$), roughly the beginning of the T-Tauri phase
\citep{Dunham_2012}, to $3\Myr$ ($\Tstr=4500K$, and $\Rstr=1.5\Rsun$).
\cite{Landry_2013} followed the viscous evolution of
  the inner $70\AU$ of the disk. Because our
goal is to probe the giant planet-forming regions,  we
  focus our study on the inner $70\AU$ of the disk.

The disk is heated passively by stellar irradiation, and actively by
viscous heating due to accretion. In order to determine the degree of
viscous heating, \cite{Landry_2013} evaluated whether MRI is present
or not in each individual grid cell by considering both Ohmic
resistivities and ambipolar diffusion. The MRI is shut down by Ohmic
resistivity on the midplane, forming a deadzone extending to about
$16\AU$. The disk has an MRI active layer near the surface, which
provides an accretion rate of $10^{-9}\Msunyr$ even when a dead zone
is present. However, unlike models with assumed uniform turbulent
efficiency [with a constant $\alpha$ parameter, where viscosity $\nu =
  \alpha c_s H$, $c_s$ is the sound speed, and $H$ is the scale
  height; \citet{Shakura_Sunyaev_1973}], heating contributed by MRI
(extracted from the energy of shear flow) only contributes slightly to
the total heat budget of the disk because the power is deposited in
the disk atmosphere at low optical depths.

\cite{Landry_2013} calculated the heating from stellar irradiation in
a $1+1$D geometry using Rosseland mean opacities. Following
\cite{Chiang_Goldreich_1997}, \cite{Landry_2013} considered stellar
radiation entering the disk at a grazing angle, heating up the disk
surface. Reprocessed stellar flux then traveled vertically from the
disk surface to the midplane. However, long-wavelength radiation
emitted by dust grains at the disk surface should travel unimpeded to
the midplane, causing heating that is not captured by using Rosseland
mean opacities. By assuming that heating propagates only vertically
from the heated surface, one also neglects the heat transfer in the
radial direction within the disk, which can be important when the
wavelength of the re-emitted light is long enough to allow radial 
propagation. One therefore underestimates the disk temperature by using a
$1+1$D geometry and Rosseland mean opacities.

The stellar irradiation cannot heat the disk enough to ionize the disk
interior, so the $1+1$D approximation with Rosseland mean opacity in
\cite{Landry_2013} was useful to calculate ionization, making the disk
evolution model tractable. However, the missing heating can be crucial
for the disk chemistry. We improved the treatment of the stellar
irradiation by including $2$D dust radiative transfer models built
with the publicly available code RADMC
\citep{Dullemond_Dominik_2004}. We then calculated the total
temperature by taking a flux-weighted sum of the accretion temperature
(T$_{acc}$) calculated from the MRI model and the equilibrium
temperature (T$_{eq}$) from stellar irradiation calculated from the
$2$D dust radiative transfer model at every point of the disk (as
demonstrated in the upper half of Fig.~\ref{fig:ModelFlowChart}.):
\begin{equation}
T^{4} = T_{eq}^{4} + T_{acc}^{4},
\label{eq:temp}
\end{equation}

The $2$D dust radiative transfer models are set up with gas density
profiles calculated by \cite{Landry_2013} and a constant gas/solid
ratio throughout the disk. The main focus of
  \cite{Landry_2013} was to study the mass transport. The disk is
  truncated at about two scale heights above the midplane. The model
  thus includes the
  majority of the disk mass, but the dust above the original disk
  surface in \cite{Landry_2013} could affect disk thermal properties
  by absorbing the stellar radiation. We therefore extrapolate the
  density profile vertically with Gaussian distributions down to a
  background density of $10^{-26} \gcmthree$ as input for the radiative
  transfer calculation.

\cite{Landry_2013} found that if the mean grain size is as small as
$0.1\um$, the MRI shuts down completely. If the MRI drives accretion
in the T-Tauri phase, some grain growth must have occurred, consistent
with the findings of \cite{Oliveira_2010}. (See \cite{Gressel_2015}
for updated models suggesting that accretion is driven by winds, not
MRI.) Evidence for rapid growth of solids in disks has been accumulating 
in the following ways:  direct evidence from  long-wavelength observations 
 \cite{Isella_CARMA_graingrowth_2010}, \cite{Guilloteau_graingrowth_2011},
 \cite{Banzatti_graingrowth_2011}, \cite{Perez_2012_AS209}, \cite{Perez_2015_CYTau_DoAr25}, 
 and \cite{Tazzari_graingrowth_2015arXiv};
   observations of forming planets in the
LkHa 15 disk by \cite{Kraus_2012_LkCa15}; and the HL Tau images from
the \cite{ALMA_HLTau_2015}, which show gaps possibly sculpted by
forming planets or resonances with planets.

\begin{figure*}
\plotone{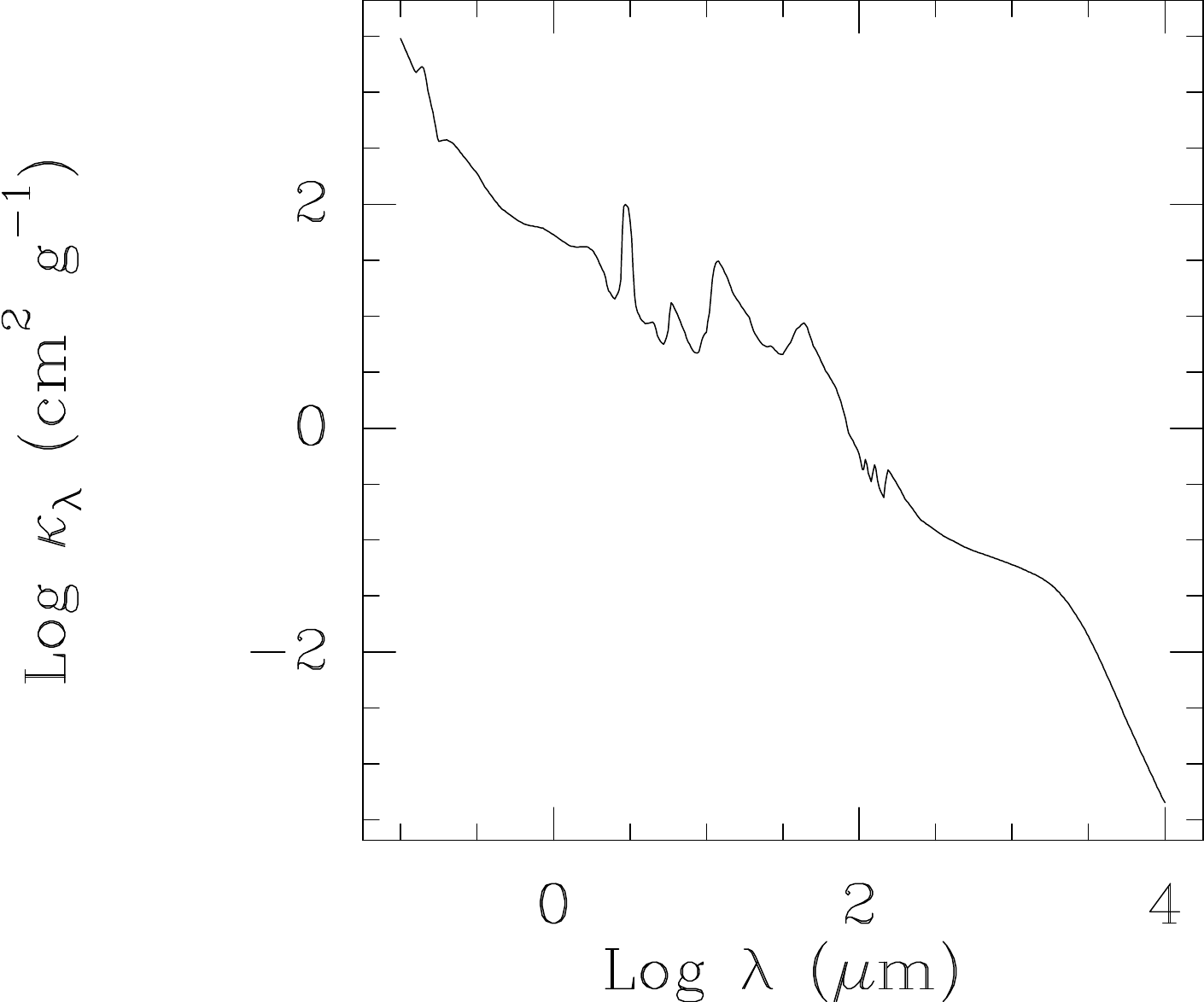}
\caption{Opacity as a function of wavelength for the 5-layer multishell
grain model of \cite{Semenov_Henning_2003}. The grain temperature
used for this plot is 155~K.
\label{fig:modeldustopacities}}
\end{figure*}

The wavelength-dependent dust opacities that we use are taken from the
website\footnote{http://www.mpia.de/homes/henning/Dust$\_$opacities/Opaciti-\\es/opacities.html},
which provides models whose Rosseland and Planck mean opacities are
described by \cite{Semenov_Henning_2003}, who updated earlier models by
\cite{Henning_Stognienko_1996} for dust in protoplanetary disks. The
available dust models include those with different assumptions about
the iron mixture in the silicates, different models of grain growth,
and different grain topologies. The dust grains are aggregated from
``sub-grains”, which themselves follow an MRN distribution extended up
to 5 \micron\ \citep{1977ApJ...217..425M,1985Icar...64..471P}.  The
dust model that we use assumes a ``normal" mix of iron [Fe/(Fe$+$Mg) =
  0.3] and that grain growth has occurred through particle cluster
aggregation (PCA in the nomenclature of Semenov), which leads to
roughly spherical grains. We use the multishell spheres topology and
take the models for temperatures up to 155 K, in which ices and
volatile organics are retained.  Figure \ref{fig:modeldustopacities}
shows the opacity per gram of gas versus wavelength for the adopted
dust properties. There are a number of resonances and a relatively
slowly declining opacity out to a wavelength of about 1.5 mm, beyond
which the opacity declines rapidly. The opacities are given per gram
of gas.

In our standard model, the dust has already grown and aggregated at
the start of disk evolution - 0.1$\Myr$ after the formation of the central star.  
We further assume that 90\% of the dust has grown to still larger
sizes (pebbles, rocks, etc.), which we no longer describe as dust.
This further growth decreases the grain surface area for chemical 
reactions and makes these solids essentially invisible even to 
millimeter wavelength observations \citep{Birnstiel_graingrowth_2011,
  Garaud_graingrowth_2013}. Consequently, we take a gas to dust ratio
of 1000, so we divide the opacities from \cite{Semenov_Henning_2003} by 10. 
In this model, there is a constant replenishment of dust by collisions
between larger objects and the size distribution of dust does not
evolve. We consider a different model for grain evolution 
as a variation of our standard
assumptions in \S \ref{sec:graingrowth}.

\begin{figure*}[ht]
\centering
\begin{tabular}{@{}cc@{}}
  \includegraphics[width=0.48\textwidth]{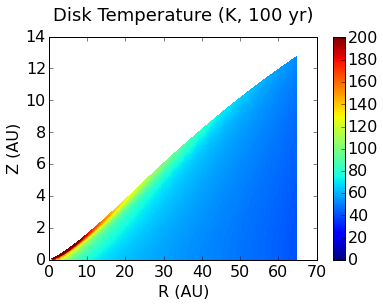} &
  \includegraphics[width=0.48\textwidth]{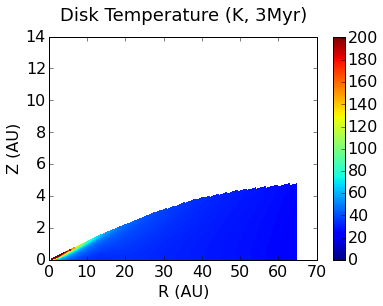} \\
  \includegraphics[width=0.48\textwidth]{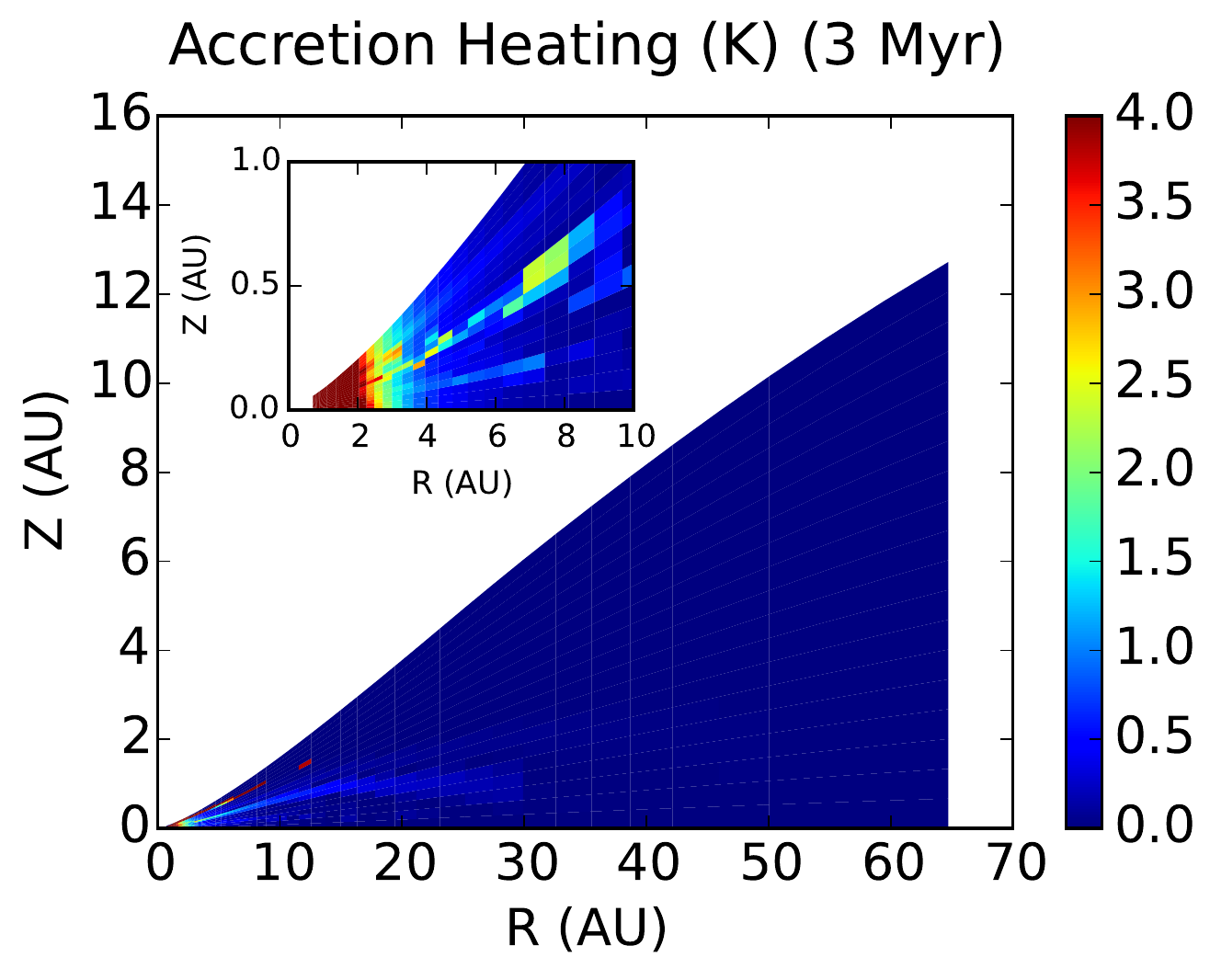} &
  \includegraphics[width=0.48\textwidth]{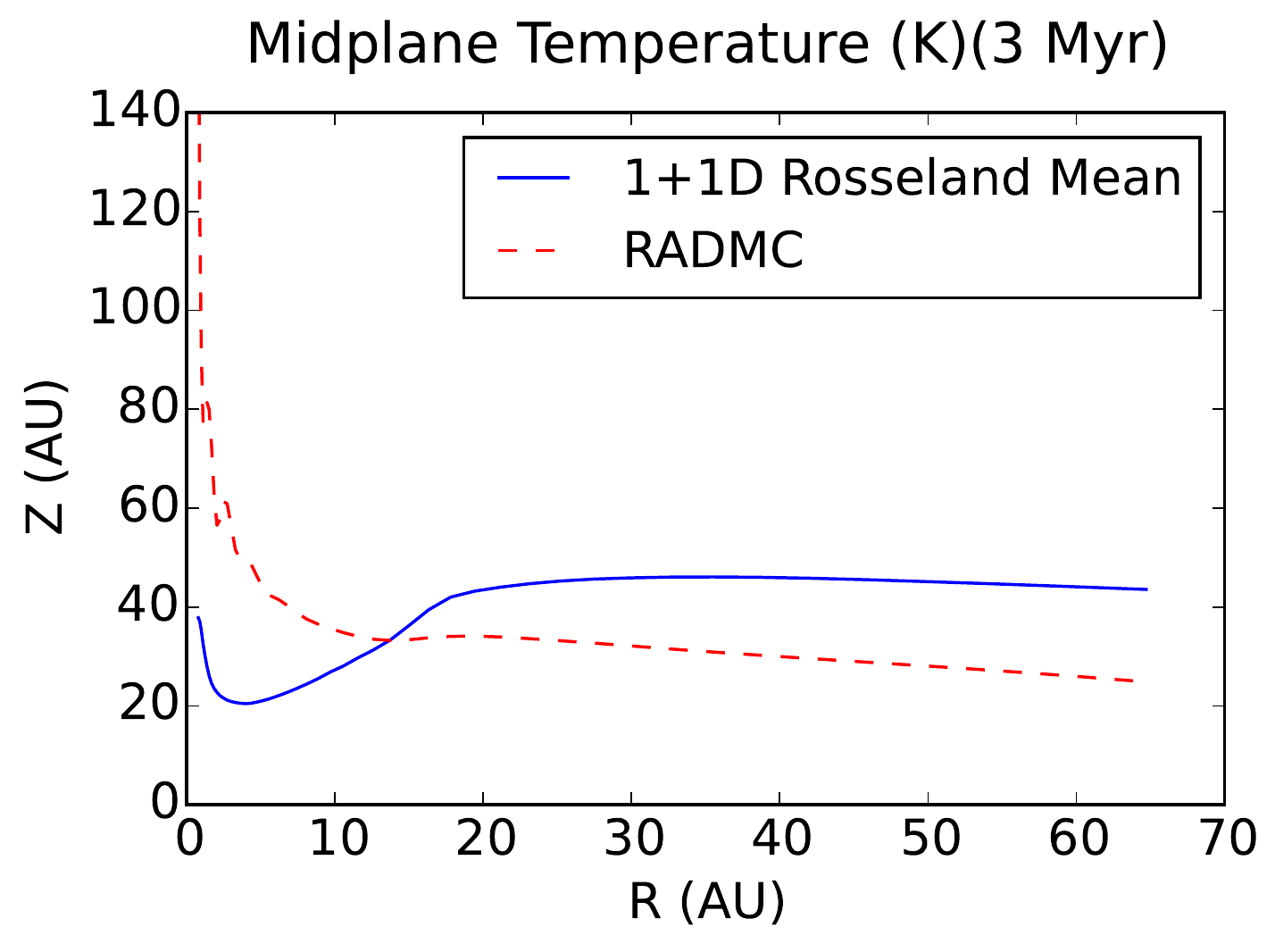} \\
  \end{tabular}
  \caption{Upper left panel: Disk temperature (100yr) as a function of
    the radial distance to the central star (R) and height above the
    disk midplane (Z); Upper right panel: Disk temperature (3Myr);
    Lower left panel: Accretion heating (T$_{acc}$) contributed by the
    MRI turbulence (3Myr); The patches on the three color coded plots are
    interpolation artifacts, although the ``arc'' of light blue at
    $\sim 1$AU above the midplane going from $5-30$ AU is real heating
    due to the MRI turbulence.  Lower right panel: A comparison of the
    midplane temperature (Teq only) computed by RADMC and the 1+1D
    model with Rosseland mean opacities (3Myr). The $1+1$D model does
    not capture the heating contributed by longer wavelength radiation
    and underestimates the disk temperature in the inner $20$AU. This
    effect is more severe at smaller radius due to larger optical
    depth, resulting in a lower temperature at smaller radius on the
    disk midplane - even though its surface temperature is higher.}
    \label{fig: TemperatureCombined}
\end{figure*}

The temperature color coded plots (at the beginning and end of the evolution)
with T$_{eq}$ computed from the $2$D dust radiative transfer model are
shown in the upper panels of Fig. \ref{fig: TemperatureCombined}. In
both cases, the temperature is the highest on the disk surface due to
the heating from the central star, and gradually decreases at larger
radius and into the disk interior. The temperature almost everywhere in
the disk decreases due to the star evolving down the Hayashi track
($\Lstr$ decreases from $12.1 \Lsun$ to $0.8 \Lsun$) and the
flattening of the disk itself (cf. the top two panels in
Fig. \ref{fig: TemperatureCombined}). The accretion heating
(T$_{acc}$) contributed by the MRI turbulence is shown in the lower
left panel of Fig. \ref{fig: TemperatureCombined}. Accretion heating
contributes less then $3 \K$ in the majority of the disk, rising to
about $4 \K$ within $2\AU$ from the central star.

We compare the midplane temperature with T$_{eq}$ computed from the
$1+1$D model after $3$Myr years of evolution in \cite{Landry_2013} and
the one computed from the $2$D dust radiative transfer model in the
lower right panel of Fig. \ref{fig: TemperatureCombined}. The
temperature is much higher in the $2$D model in the inner $20\AU$. The
$1+1$D model misses heating in the inner part of the disk where the
optical depth is larger, and therefore produces artificially lower
temperature on the midplane closer to the central star. For example, the
Rosseland mean optical depth is $\sim 12$ at $5$ AU, but $\sim 4$ at
$15$ AU, so $5$ AU is more severely affected---even though its surface
temperature is higher. Moreover, once the temperature is down to $\sim
50$ K, the black-body radiation re-emitted from dust peaks at
far-infrared wavelengths. The mean free path is a significant fraction
of the scale height and the assumptions used to justify using
Rosseland mean opacity breaks down.

Our treatment of temperature is not ideal in the sense that the
vertical density structure and the temperature structure are not
calculated consistently. However, as an experiment we artificially
increased the scale height everywhere in the disk by $20$ percent and
did not find significant changes to the midplane temperature. We also
assume the gas and dust are well mixed and have the same temperature
in this paper, which is a valid assumption except for the very surface
of the disk. Since the purpose of this project is to find an optical
depth in the vertical direction and we mainly focus on the disk
midplane where most of the disk mass resides, the combined accretion
and passive heating models, assuming the same temperature for the gas
and dust, meet our needs.

In the next sections, we describe our standard model and results.
Then we describe two variations in the chemistry and one variation in
the model of dust evolution.

\section{Chemical model}   

The disk temperature and density evolution described in
Section~\ref{sec:thermodynamic} allow us to compute the chemical evolution of
the disk. 

We also need initial chemical abundances, which we
obtain from a simplified model of the molecular cloud.

We construct chemical evolution models including C, H, O, N,
and different C and O isotopes based on the UMIST database RATE06
\citep{Woodall_UMIST_2007}, and follow the chemistry of $588$ species,
$414$ gas-phase and $174$ ices. The reaction network contains $13116$
reactions, including gas-phase reactions, grain-surface reactions,
freezeout, thermal desorption, and reactions triggered by UV, X-rays
and cosmic rays, such as isotope-selective photodissociation. The
carbon isotopic chemistry network was developed by
\citet{Woods_Willacy_2009} and was extended to include oxygen isotopes
for this work.

We calculate the chemical evolution at each grid point
independently. By doing this, we assume that the radial and vertical
motions of gas and dust are slow compared to chemical reaction
timescales, and that mixing is not important in determining the
chemistry. This is a valid assumption for gas-grain reactions or grain
surface reactions due to their short reaction timescale, but might not
work as well for gas-phase reactions. However, given that turbulence
in the inner $15$ AU is restricted to the disk surface, we do not
expect vertical mixing to contribute much to disk chemistry.

In order to calculate the chemical evolution over a $3\Myr$ lifetime,
we use an extension of the computationally efficient rate-equation
method to compute reaction rates for grain-surface reactions under the
``mean field'' approximation (as used by
\citealt{Dodson-Robinson_2009}). Rate equations neglect the stochastic
variation of abundances on different grains and are not appropriate
when the number of reacting particles per grain is less than one. We
modify our reaction rates following the method of
\cite{Garrod_Pauly_2011}, which considers the competition between
thermal hopping of mobile species and reactions on the grain
surfaces. We assume that only atoms and simple hydrides are mobile on the
grains. Our treatment of gas phase reactions, freezeout, and thermal
desorption is similar to that of \cite{Dodson-Robinson_2009}. After
describing the molecular cloud preprocessing model in section
\ref{cloud}, we focus the rest of this section on the treatment of
photochemical and cosmic-ray reactions, which are new to this work.

\subsection{Preprocessing in the molecular cloud}
\label{cloud}

We model the chemical evolution in the molecular cloud stage to
derive input abundances for the protoplanetary disk models. 

The input abundances are listed in Table~\ref{table:MC_input}
\citep{Graedel_1982_MCinput}. Throughout this paper, abundances are
presented as the number density normalized to the number density of
hydrogen nuclei, n$_{x}$/(n$_{\rm H}+2\rm n_{\rm H_{2}}$). 

The molecular cloud model is run for 1 Myr with a density of $2\times 10^{4}$
cm$^{-3}$, at a temperature of $10$K and with a visual extinction of
$10$ magnitudes. 

The abundances at the end of the molecular cloud phase
are given in Table \ref{table:MC_c2d}.
The $^{12}$C/$^{13}$C is $\sim$ $46$ in CO gas, $\sim$
  $63$ in CO ice and $\sim$ $58$ in CO$_2$ ice at the end of the
  molecular cloud model -- all lower than the initial elemental
  $^{12}$C/$^{13}$C ratio.  The reduction is caused by the
  preferential formation of heavier isotopomers of CO arising from the
  small energy different in the ion exchange reaction.

\begin{equation}
^{13}\mathrm{C}^{+}+{\rm CO}\leftrightarrow ^{13}\mathrm{CO} + \mathrm{C}^{+}+35K.
\label{eq:CO}
\end{equation}

$^{13}$CO has a lower ground-state vibrational energy
  due to its slightly larger mass compared to CO, and is therefore
  more energetically favorable.  At $10$ K Equation \ref{eq:CO}
  dominates the carbon fractionation, leading to lower
  $^{12}$CO/$^{13}$CO than the elemental $^{12}$C/$^{13}$C ratio (see
  also \citealt{Visser_CO_2009}, \citealt{Langer_C}).  In other molecules, such
  as CH$_{4}$ ice, the opposite effect is seen, with enhanced abundances
  of condensed $^{12}$C leading to higher $^{12}$CH$_{4}$/$^{13}$CH$_{4}$ 
  ratios in the ice.

\subsection{UV ionization, photodissociation and photodesorption}
\label{photochem}

In addition to ionizing and dissociating gas phase species, UV photons
also desorb molecules from grain mantles.  The UV flux generated by a
young star has been observed by \cite{France_UV_2014} who found a
median value of $1000$ ISRF (1 ISRF = 1.6 $\times$ 10$^{-3}$
ergs$^{-1}$ cm$^{-2}$; \cite{Habing_ISRF_1968}) at $100$ AU.
Theoretical studies of \cite{ACP_Photoevaporation_2006} suggest a UV
flux of $\sim$ 50 ISRF at $100$ AU is required to drive
photoevaporation of the disk.  We choose a UV flux between these two
values of $500$ ISRF at $100$ AU.

The UV field is attenuated by the disk and the resulting visual extinction can be related to the column density by 
\begin{equation}
A_{V} (r,z) = 5.2\times{10}^{-22}\times\frac{f_{H}\Sigma_{r}(r,z)}{2m_{H}},
\label{eq:av}
\end{equation}
where $f_{H}=0.735$ is the mass fraction of hydrogen  in the Sun \citep{Grevesse_Sauval_1998}, and $\Sigma_r(r,z)$ is the
horizontal column density integrated from the inner edge of the disk.

For photodesorption we adopt the reaction rate of \citet{Hollenbach_2009}

\begin{equation}
k(r, z) = F(r,z) \times{10}^{8}\times(\pi{a}^2){\times}Y,
\label{eq:photdrate}
\end{equation}

where F(r,z) is the UV flux in ISRF, $\pi a^2$ is the grain
cross-section and Y is the photodesorption yield. The average grain
radius $a$ is taken to be $1$ \micron, and $Y$ is assumed to be 10$^{-3}$
for all molecules (based on laboratory measurements of H$_2$O
photodesorption by Ly$\alpha$ photons (Westley et al. 1995a, b).

For the self-shielding of CO and H$_2$ we use the method of
\cite{vanDishoeck_Black_1988} and \cite{Lee_H2SS_1996} respectively.

\subsection{X-ray photoionization}

We follow the X-ray ionization prescription of
\cite{Bai_Goodman_2009}, using a value of the ionization rate for
direct absorption of X-rays, $\zeta_{abs}$ = 6 x 10$^{-12}$ s$^{-1}$,
and for scattered photons, $\zeta_{sca}$ = 10$^{-15}$ s$^{-1}$
\citep{Igea_Glassgold_1999}.  We assume a stellar X-ray luminosity
$L_X$ = 20 $\times$ 10$^{29}$ erg s$^{-1}$ based on the median
observed value from a Chandra survey of solar mass young stars in the
Orion Nebula \citep{Garmire_xray_2000}.  Similar values have also been
observed in the Taurus-Auriga complex by \cite{Telleschi_Xray_2007}
and \cite{Robrade_Xray_2014}. The X-ray ionization rate is assumed to
be the same for all reactions because of the lack of laboratory
measurements.

\subsection{Cosmic Rays}

Cosmic rays can ionize and dissociate molecules in the gas directly,
or by producing secondary photons \citep{Gredel_CRPHOT_1989}.  The
intensity of cosmic rays decreases exponentially with a characteristic
attenuation depth of 96 g cm$^{-2}$ \citep{Umebayashi_CR_1981,
  Umebayashi_CR_2009}.  The large penetration depth means that our
model disk is transparent to cosmic rays outside of 4 AU at the
beginning of the T Tauri phase and outside of $3$ AU at the end of $3
\Myr$ evolution.  Cosmic ray reaction rates are taken from UMIST06
with an assumed cosmic ray ionization rate of 1.3 $\times$ 10$^{-17}$ s$^{-1}$.

\subsection{A summary of ionization processes} 

\begin{figure*}[ht]
\centering
\begin{tabular}{@{}cc@{}}
\includegraphics[width=0.48\textwidth]{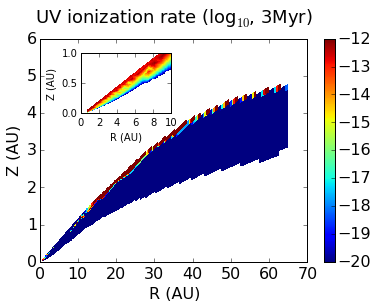} &
 \includegraphics[width=0.48\textwidth]{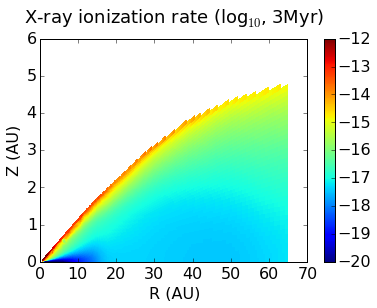} \\
 \includegraphics[width=0.48\textwidth]{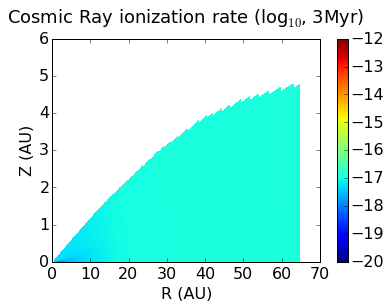} &
 \includegraphics[width=0.48\textwidth]{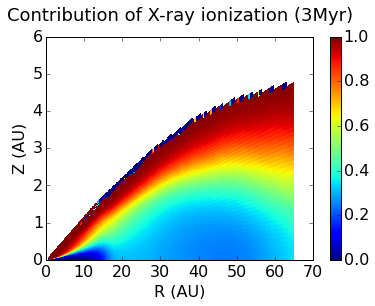} \\
  \end{tabular}
  \caption{Estimated ionization rates contributed by various
    mechanisms at the end of our disk evolution.  Upper left panel: UV
    ionization rate (an order of magnitude estimation for the H$_{2}$O
    molecule, the ionization rate in the disk interior is lower than
    $10^{-20} s^{-1}$, and therefore does not show up in the plot.);
    Upper right panel: X-ray ionization; Lower left panel: cosmic ray
    ionization (an order of magnitude estimation considering only the
    H$_{2}$ molecule); Lower right panel: fraction contributed by
    X-ray ionization. X-ray ionization dominates at z/r $>0.1$, and
    cosmic rays account for most of the ionization for z/r $<0.1$,
    where the UV and X-ray radiation from the central star are
    sufficiently attenuated.}
    \label{fig: Ionization}
\end{figure*}

Ionization rates contributed by the above three ionizing sources are
shown in Fig.~\ref{fig: Ionization}. The upper left panel shows an
order-of-magnitude estimate of the UV ionization rate based on the
H$_{2}$O molecule, which is the molecule most commonly ionized by
UV. The ionization rate decreases very rapidly due to the small
penetration depth of UV photons, and it is negligible except for the
very surface layer of the disk. X-rays are able to reach most of the
disk except for the inner $10\AU$ near the disk midplane (shown in the
upper right panel in Fig.~\ref{fig: Ionization}). They provide
ionization rates of ${10}^{-14}$~s$^{-1}$ to ${10}^{-13}$~s$^{-1}$
near the disk surface, and a modest ionization rate around
${10}^{-17}$~s$^{-1}$ in the disk interior due to scattering. In the
lower left panel, we show the cosmic-ray ionization rate of H$_{2}$,
the most common cosmic-ray reaction partner. Cosmic rays provide a
steady ionization rate around ${10}^{-17}$~s$^{-1}$ throughout the
disk. The fractional contribution of X-ray ionization is plotted in
the lower right panel of Fig.~\ref{fig: Ionization}. Roughly speaking,
X-rays dominate the ionization rate above one scale height of the disk
where the vertical column density is less than a few g~cm$^{-2}$, and
the cosmic ray ionization dominates the disk interior within one scale
height of the midplane. Cosmic ray-induced photons are abundant enough
to cause ionization and photodissociation throughout the disk due to
the efficient penetration of cosmic rays. Ionization rates due to
cosmic ray-induced photons are different for each reactant and are not
shown in Fig.~\ref{fig: Ionization}. However, cosmic ray-induced
photons are important contributors to the chemical evolution.

\cite{Cleeves_2013_CR} showed that stellar winds can power a
``T-Tauriosphere’’ that shields the disk from external cosmic rays,
leading to cosmic ray ionization rates of ${10}^{-18}$~s$^{-1}$ or
lower. A decreased cosmic-ray flux would reduce the rates of reactions
with both cosmic rays and cosmic ray-induced photons. On the other
hand, the decay of shortlived radionuclides (SLRs) such as $^{26}$Al
can provide an ionization rate on the order of ${10}^{-19}$~s$^{-1}$
to ${10}^{-18}$~s$^{-1}$ \citep{Cleeves_2013_SLR}, which may help to
drive the chemistry in the disk interior. Since our chemical evolution
models are computationally expensive, we do not explore different
values of ionization rates.

\section{Chemical model results}

The results of our chemical model give the abundances of CO
isotopologues that we need for the optical depth calculation. First, we
discuss the ice line locations for different volatiles (Section
\ref{sec: Iceline_locations}). Knowing the ice line locations is
important for interpreting observed radial abundance gradients. Second,
we discuss the active reaction network involving carbon-bearing
molecules, which causes the CO abundance to change with both radius and
time (Section \ref{sec:CO}). In Section \ref{sec: CO2COM}, we
demonstrate that dissociation of CO and subsequent formation of complex
organic molecules (COM) can produce CO depletion that mimics an ice
line. In Section \ref{sec:COM_uncertainty}, we assess how computational
limits on the number of species in our reaction network impact our COM
abundances. We discuss our model's dependence on initial
cloud abundances, reaction rates, and grain evolution in Section
\ref{sec:chem_dependence}. We show that despite the uncertainty in the
exact end product of COM formation, the formation of ices as carbon
sinks on grain surfaces is robust, and leads to the depletion of CO in
the gas phase. In order to connect the abundances of rare CO
isotopologues with the overall disk mass, we examine the carbon
fractionation in CO and other major carbon-bearing molecules in section
\ref{sec:CO_frac}.

\subsection{Locations of ice lines}
\label{sec: Iceline_locations} 

Due to efficient heating from the central star, CO and CH$_4$ do not
freeze out in our modeled region---the inner $70\AU$ of the disk---at
any time in our 3~Myr of evolution (though disks with different grain
properties or less luminous host stars may be colder than our model disk).
Using CO isotopologues to estimate planet-forming mass therefore does
not require correcting for CO freezeout, at least in disks surrounding
proto-Sunlike stars. However, as we will see in Section \ref{sec:
CO2COM}, freezeout of other organic molecules affects the gaseous CO
abundance. H$_2$O and CO$_2$ freeze out at $1.5\AU$ and
$18\AU$ at the beginning of disk evolution, and their condensation
fronts move inward to $1\AU$ and $2\AU$ as the disk becomes cooler. We
see ices of hydrocarbons and other carbon-bearing molecules formed from
CO  after a few hundred thousand years of disk
evolution. The abundance of an ice depends on both condensation
temperature and formation pathway: a molecule may have been chemically
destroyed and simply not be present to freeze out. Our discussion of ice
lines focuses only on locations where the relevant molecule exists. We
defer the discussion of molecule formation and destruction to the next
two sections.

The binding energies and the locations where different ice species exist
after 3~Myr of disk evolution are shown in Table
~\ref{table:IceLine_BindEng}. Species with larger binding energy can
form stronger bonds with the grain surface, and therefore freeze out at
higher temperature. In our model, species with binding energy larger
than $E_B / k = 2.5 \times 10^3$~K can freeze out well within the giant
planet-forming region at $\la 15$ AU
\citep{Tsiganis_NICE_2005,Thommes_2002}; species with binding energy of
$2.1 \times 10^3$~K can only freeze out in the outer part of the disk or
at later stage of evolution when the temperature is lower, and species
with binding energy lower than $10^3$~K---including CO---do not freeze
out.

The condensation of volatiles is important for the growth of planets due
to the increase of available solid surface density
\citep{Dodson-Robinson_2010_UN, AliDib_UN_formation_2014}, and is
crucial for determining the chemical composition of giant planets
\citep{Oberg_2011_CtoO}. Moreover, the effect of radial
drift and gas accretion may cause further movement of ice lines and
changes in chemical composition \citep{AliDib_transportation_2014,
Piso_CtoO_snowline_2015}. However, the above studies assumed CO to be a
major carrier of carbon and oxygen, which may not be true throughout
the disk evolution. Our results indicate that it is important to take
into account the possibility that CO ice is not a mass source for giant
planets in some systems. Instead, CO$_2$, hydrocarbons, and
methanol may be the major carbon ice reservoirs that contribute
to planet growth.

Furthermore, because of the unprecedented sensitivity and resolving
power of (sub)mm interferometers such as the Atacama Large
Millimeter/submillimeter Array (ALMA), the location of the CO ice line
has been considered an important temperature tracer of giant
planet-forming regions (\citealt{Qi_snowline_HD163296,Qi_N2H1_2013}).
\cite{Qi_N2H1_2013} use observations of N$_2$H$^+$, which CO destroys,
to infer a CO snow line radius of $\sim 30$~AU in the disk surrounding
TW~Hya. If some combinations of grain size, star luminosity and
UV/X-ray/cosmic-ray flux push CO ice lines outside of ~$70$ AU, CO
freezeout may not trace the region of giant planet formation. Even
TW~Hya, at 0.8~$M_{\odot}$ and 10~Myr, is luminous enough to push its
disk's CO snow line beyond the likely formation locations of Uranus and
Neptune in the solar nebula.

\subsection{Time evolution and spatial distribution of CO gas}
\label{sec:CO} 

Inferring disk structure based on CO line intensity is complicated by
the fact that the abundance of CO evolves significantly as a function
of time. In the inner $20\AU$ from the central star when the
temperature is high enough for CO$_{2}$ to be in the gas phase,
CO$_{2}$ is dissociated into CO~+~O by cosmic ray-induced photons at a
rate around $10^{-18}$ per second, which leads to an increase of CO
abundance on a $1\Myr$ time scale. On the other hand, although CO does
not freeze out in our modeled region within $70\AU$, the abundance of
CO drops beyond $15\AU$ because carbon is tied up in hydrocarbons,
methanol, and ketene (complex organic molecules or COMs), mimicking
the effect of CO freezeout.  Abundant carbon-bearing species include
C$_2$H$_2$ (acetylene), C$_2$H$_5$, CH$_3$CHO (acetaldehyde), CH$_3$OH
(methanol), and H$_2$CCO (ketene).  The CO depletion
rate is driven by the ionization rate of He$^{+}$ from X-rays and
cosmic rays, as He$^+$ drives the breakup of CO (section \ref{sec:
  CO2COM}). The depletion of CO beyond $15\AU$ happens on a $1\Myr$
time scale. As a result, the CO abundance depends strongly both on
location and time. As long as the disk is opaque to the UV photons
that dissociate CO, the column density of CO can increase in the inner
part of the disk even as the overall disk mass is decreasing.

\begin{figure*}[ht]
\centering
\begin{tabular}{@{}cc@{}}
  \includegraphics[width=0.48\textwidth]{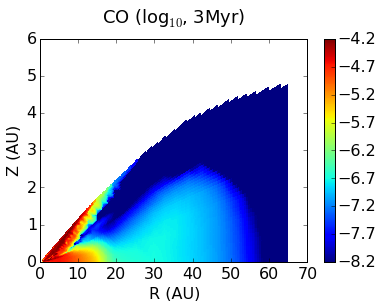} &
  \includegraphics[width=0.48\textwidth]{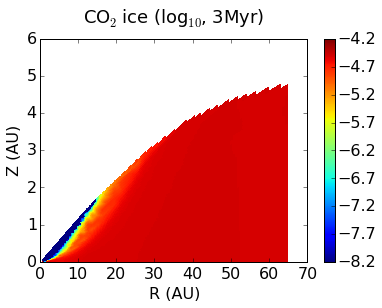} \\
  \includegraphics[width=0.48\textwidth]{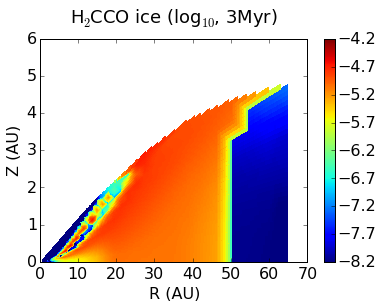} &
  \includegraphics[width=0.48\textwidth]{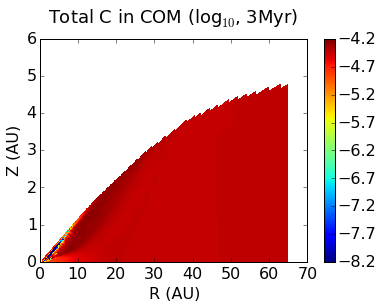} \\
  \end{tabular}
  \caption{Fractional abundances at the end of the $3$Myr evolution
    (the number density with respect to the number density of hydrogen
    nuclei, n$_{\rm H}+2\rm n_{\rm H_{2}}$). The color scale is in
    logarithm.  Upper left: CO exists in a large abundance for
    r$<15$AU; Upper right: CO$_{2}$ ice exists in most part of the
    disk where the temperature is low enough for it to stay on grain
    surfaces; Lower left: H$_2$CCO (ketene) ice is the major carbon
    sink beyond $15$ AU from the central star; Lower right: Other
    complicated organic molecules such as C$_{2}$H$_{x}$, and
    CH$_3$CHO (acetaldehyde) exist in a layer between $10-30$ AU,
    closer to the disk surface.  }
    \label{fig: Chem_combined_1}
\end{figure*}

\begin{figure*}[ht]
\centering
\begin{tabular}{@{}cc@{}}
  \includegraphics[width=0.48\textwidth]{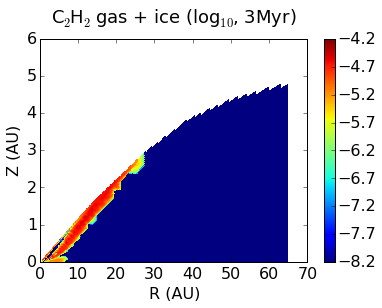} &
  \includegraphics[width=0.48\textwidth]{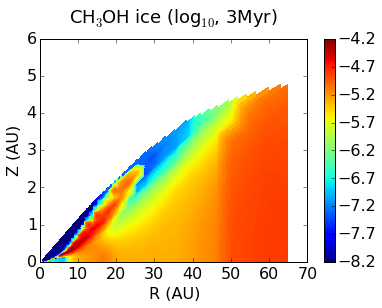} \\
  \includegraphics[width=0.48\textwidth]{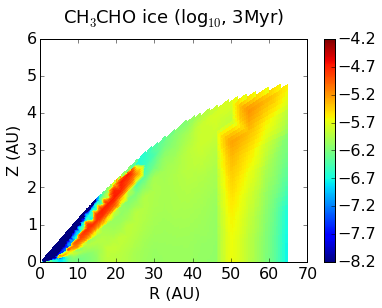} &
    \includegraphics[width=0.48\textwidth]{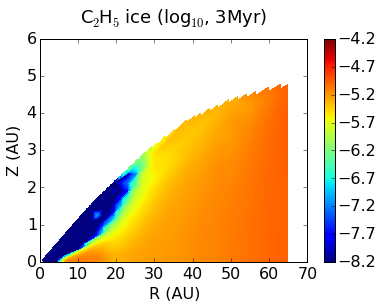} \\
  \end{tabular}
  \caption{Fractional abundances continued. The figure setup is the same as in Fig. \ref{fig: Chem_combined_1}.   
  Upper left: C$_2$H$_2$ is abundant where the temperature is too high for any carbon-bearing ices to freeze out;
  Upper right and lower left: CH$_{3}$OH and CH$_3$CHO serve as the carbon sinks where the temperature is high enough to evaporate H$_2$CCO (ketene) ice;
  Lower right: C$_{2}$H$_{5}$ is able to form where the temperature is low enough for C$_2$H$_3$ to stay on grain surface and hydrogenate.
  }
    \label{fig: Chem_combined_2}
\end{figure*}

Color-coded plots of abundances of major carbon-bearing molecules are shown
in Fig. \ref{fig: Chem_combined_1} and Fig. \ref{fig: Chem_combined_2}.
We see abundant CO within $15$ AU from the central star, and abundant
CO$_{2}$ ice in parts of the disk where the temperature is low enough
for it to stay on the grain surface. H$_2$CCO (ketene) ice is found to
be the major carbon sink between $5$ AU and $45$AU from
the central star in our model. Other COMs such as C$_{2}$H$_{x}$,
CH$_3$OH (methanol) and CH$_3$CHO (acetaldehyde) exist in a layer
between $2-30$ AU, closer to the disk surface, or the whole disk thickness beyond $40$ AU from
the central star (lower right panel of Fig. \ref{fig: Chem_combined_1}). Again, the change of abundance can be very gradual.
Here we quote the boundary location where $10\%$ of carbon is stored in
corresponding species. At the end of our $3$ Myr disk evolution,
$13.6\%$ of available carbon is contained in CO gas,
$36.2\%$ in CO$_2$ ice, and $44.8\%$ in complex organic molecules. The
above values are integrated over the entire disk, weighted by the disk
mass in different locations.  A detailed breakdown of abundances of
various species can be found in the first column of Table
~\ref{table:comp_output}.
We present abundances in the
form of percentage of carbon contained in each species, integrated
over the entire disk, and weighted by disk mass in different
locations.

\begin{figure*}[ht]
\centering
\begin{tabular}{@{}c@{}}
  \includegraphics[width=0.60\textwidth]{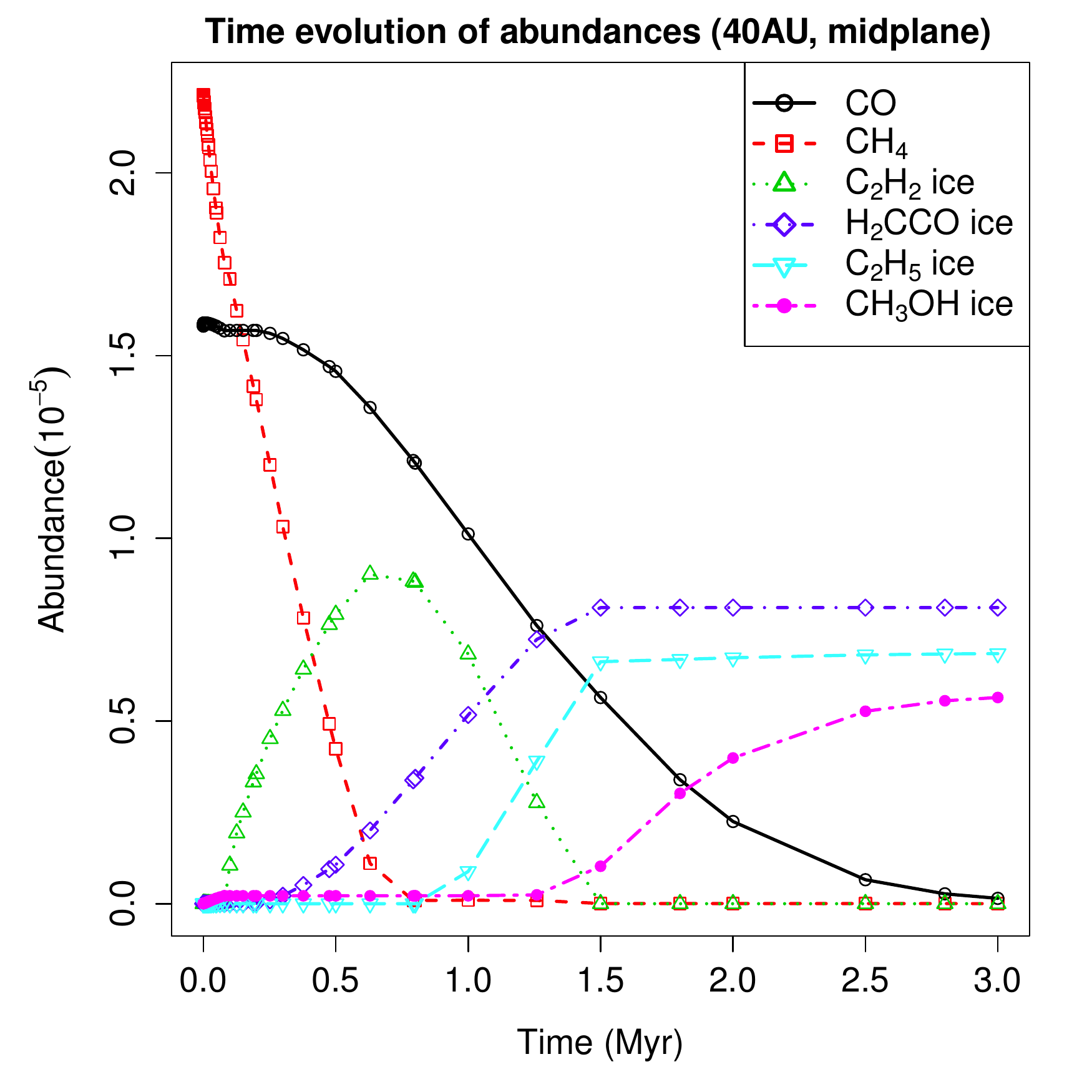} \\
  \includegraphics[width=0.60\textwidth]{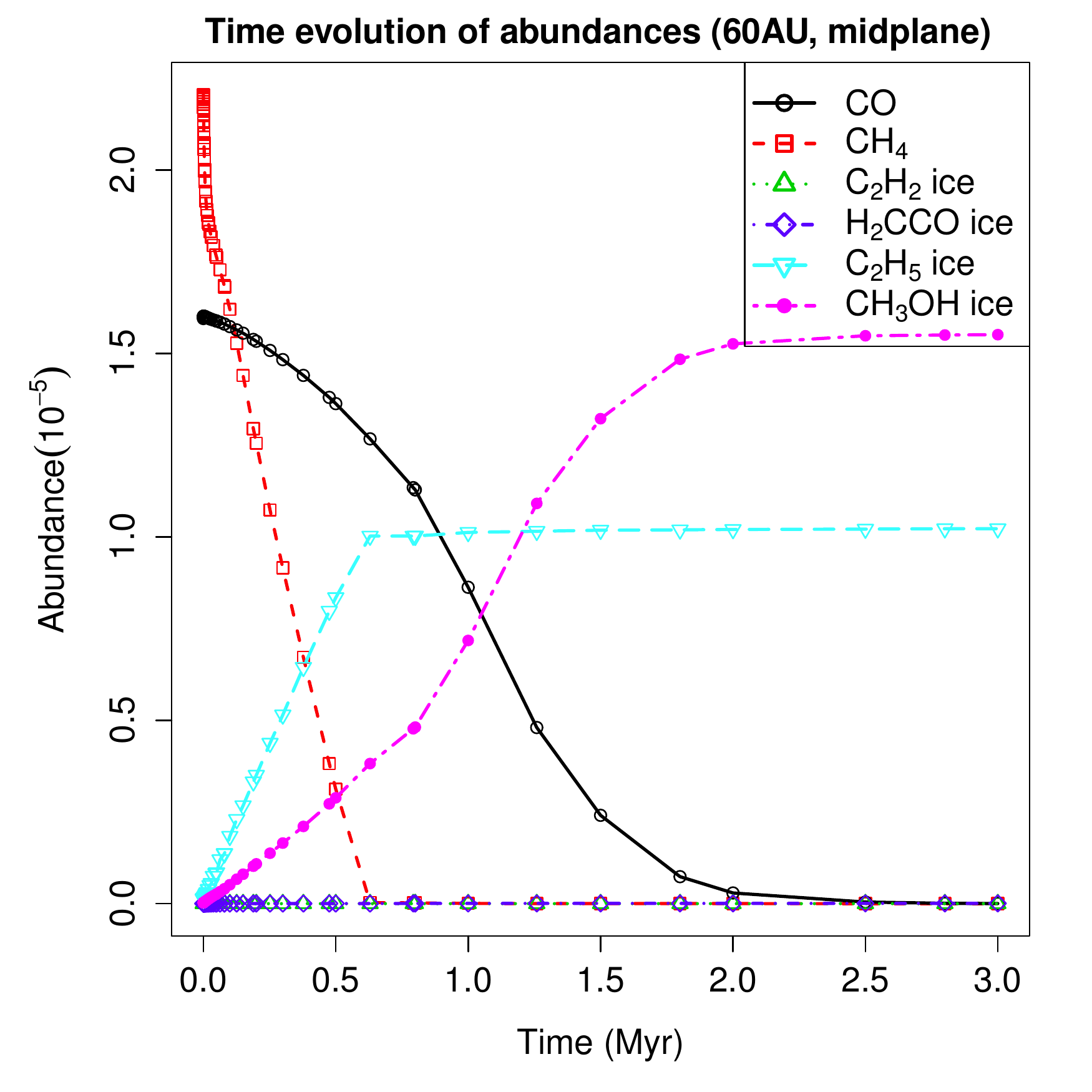} \\
  \end{tabular}
  \caption{Abundances of major carbon-bearing molecules as functions
    of time. Upper panel: $38\AU$ on the disk midplane; lower panel:
    $60\AU$ on the disk midplane. Points show the values of actual data points in our models.}
    \label{fig: time_evolution}
\end{figure*}

The time evolution of the CO/H$_2$ abundance ratio (Fig. \ref{fig: time_evolution}) over 3 Myr is substantial. 
The variation is not well-represented
by a step function as in simple freeze-out models. 
Models of CO abundance versus radius appropriate to the star's
age must be used to compute available planet-forming mass
from CO isotopologue line intensities. 

While this paper does not contain
a parameter study of CO/H$_2$ abundance ratio as a function of disk mass
and UV/X-ray/cosmic-ray flux, we suggest that flux in high-velocity line
wings (produced by CO gas near the star) may increase as the star ages,
even despite an overall reduction in disk mass. Furthermore, depletion
of CO gas in the outer disk does not necessarily mean there is CO frozen
on grain surfaces, an idea we explore further in the next section.

\subsection{CO depletion due to the formation of complex organic molecules}
\label{sec: CO2COM}

Here we investigate the causes of CO depletion in the outer disk, beyond
$15$ AU. As we can see in Fig. \ref{fig: Chem_combined_1}
and Fig. \ref{fig: Chem_combined_2}, H$_2$CCO (ketene) ice is the
dominant form of carbon within $45\AU$ from the central star in the
CO-depleted region. Beyond 45~AU, CH$_3$OH, C$_{2}$H$_{x}$, and CH$_3$CHO
ice are more abundant. To demonstrate the chemical evolution in these
two different locations, we plot the abundances of major carbon-bearing
molecules as a function of time in Fig. \ref{fig: time_evolution} for two
locations of the disk: $38\AU$ on the midplane and $60\AU$ on the
midplane.

At $38\AU$ on the midplane, the depletion of CO happens
  in three stages. In stage 1, the first $0.6\Myr$, CH$_{4}$ and CO
  react to form C$_2$H$_2$ through two different paths (Path $1$ and 
  Path $2$, see the next paragraph for details), both of which depend on
  the existence of CH$_{4}$ gas. This leads to a net
  destruction of CO and methane and increase in C$_2$H$_2$ abundance.
  C$_2$H$_2$ freezes out on grain surfaces, but because of the low
  binding energy of C$_2$H$_{3}$, C$_2$H$_2$ can not efficiently
  hydrogenate until the temperature decreases. The formation of more
  complex organic molecules such as H$_2$CCO and C$_2$H$_5$ happens roughly between $0.6$ to $1.5\Myr$
  (Stage 2) and only after the reactions in Stage 1 have already begun
  to deplete CO.  CH$_3$OH ice forms in Stage 3 after the formation of
  H$_2$CCO and C$_2$H$_5$ ices.  The CO gas abundance continues to
  decrease through all three stages. At $60\AU$ on the
    midplane where temperature is lower, C$_2$H$_{3}$ can stay on the
    grain surface and hydrogenate to C$_2$H$_{5}$ early in the disk
    evolution. We see rapid C$_2$H$_{5}$ formation in the first
    $0.6\Myr$, and the formation of hydrocarbons stops after CH$_{4}$
    is depleted. Instead, the net transfer of carbon is from CO to
  CH$_3$OH.

The formation of COM primarily follows two paths:
  reactions between CH$_x$ radicals (Path $1$), and ion-neutral reactions between
  C$^{+}$ and CH$_x$ (Path $2$). The first pathway starts from CH$_4$
  dissociation by secondary photons (h$\nu$) generated by cosmic ray
  ionization:

\begin{equation}
\mathrm{h}\nu+{\rm CH}_{4} \to \mathrm{C}\mathrm{H}_{2} + \mathrm{H}_{2}.
\label{eq:ch4crphot}
\end{equation}

CH$_{2}$ then reacts to form CH and subsequently combines with CH$_{4}$ to form C$_{2}$H$_{4}$.

\begin{align}
\mathrm{H}+{\rm CH}_{2} &\to \mathrm{CH} + \mathrm{H}_{2}\\
\mathrm{CH}+{\rm CH}_{4} &\to \mathrm{C}_{2}\mathrm{H}_{4} + \mathrm{H}.
\label{eq:c2h4}
\end{align}

Similar reactions involving other CH$_x$ radicals include:
\begin{align}
\mathrm{CH}_{2}+{\rm CH}_{2} &\to \mathrm{C}_{2}\mathrm{H}_{3} + \mathrm{H}\\
\mathrm{C}+{\rm CH}_{3} &\to \mathrm{C}_{2}\mathrm{H}_{2} + \mathrm{H}.
\label{eq:c2formation_radicals}
\end{align}

In a relatively high temperature and hydrogen rich environment, C$_2$H$_3$ and C$_2$H$_4$ formed in above reactions quickly react into a more stable form - C$_2$H$_2$.

\begin{align}
\mathrm{C}_{2}\mathrm{H}_{3}^{+}+\mathrm{C}_{2}\mathrm{H}_{4} &\to \mathrm{C}_{2}\mathrm{H}_{5}^{+} + \mathrm{C}_{2}\mathrm{H}_{2}\\
\mathrm{H}+\mathrm{C}_{2}\mathrm{H}_{3} &\to \mathrm{C}_{2}\mathrm{H}_{2} + \mathrm{H}_{2}.
\label{eq:backtoc2h2}
\end{align}

\begin{figure*}[ht]
\centering
\includegraphics[width=\textwidth]{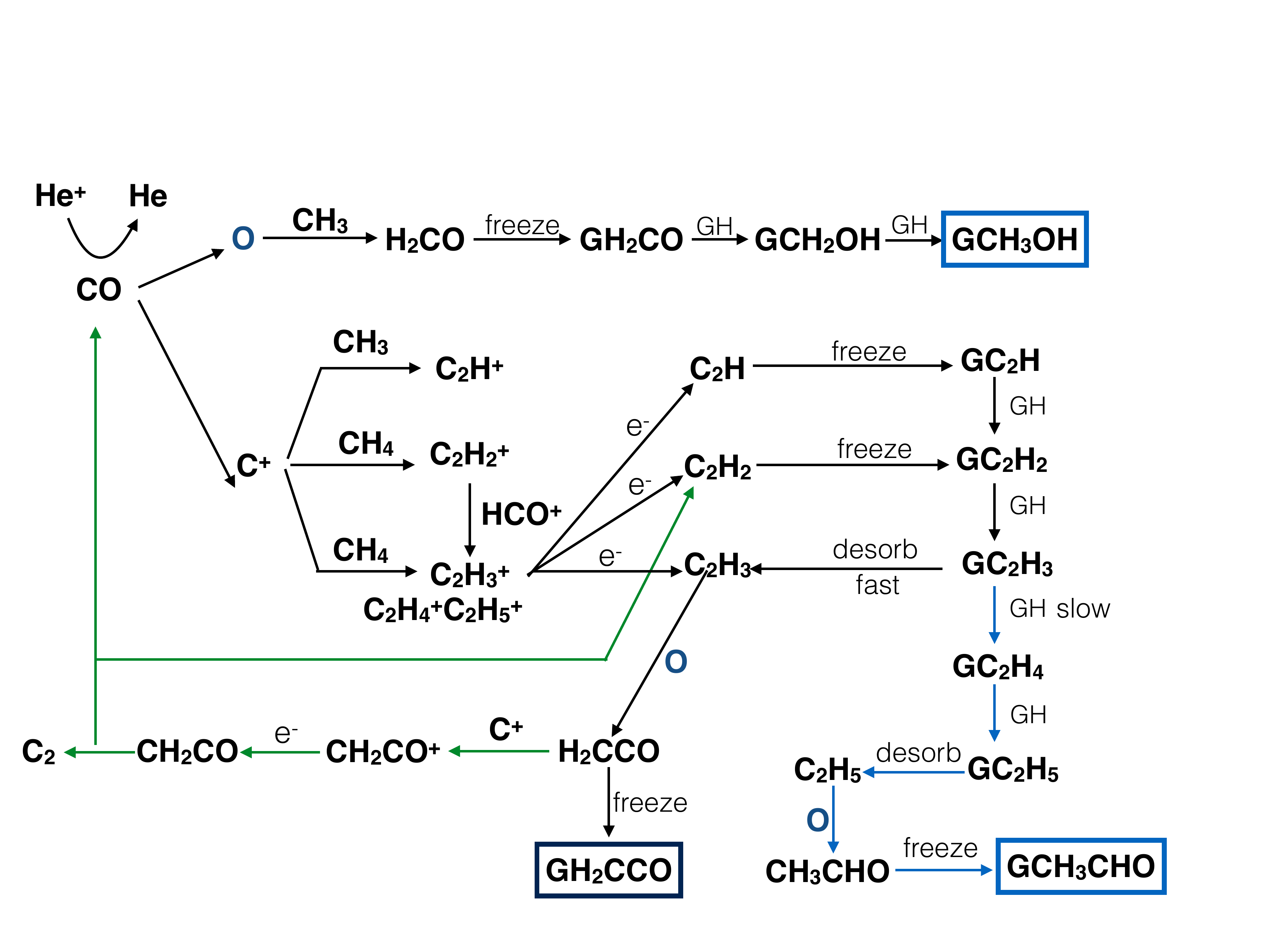}
\caption{Reaction network for major carbon-bearing species. The
  species with boxes drawn around them are sinks, and letter G denotes
  species that are frozen out on grain surfaces. Starting from the
  upper left side the the figure: (1) The dissociation of CO is
  initiated by He$^{+}$; (2) The topmost pathway shows methanol
  formation in relatively hot regions of the disk, where H$_{2}$CCO
  molecules are not able to stay on the the grain surface.  (3) The
  blue lines represent processes that move carbon from C$_2$H$_3$ to
  GCH$_3$CHO (acetaldehyde); (4) The green lines trace a path for
  removing carbon from ketene in warm parts of the disk; To summarize,
  C$_2$H$_2$ gas exists where the temperature is too high for icy
  carbon sinks to form. GCH3CHO and GCH3OH are the carbon sink in warm
  regions, GH2CCO in the majority part of the disk where the
  temperature is lower.  }
\label{fig: carbon_full}
\end{figure*}

While the major carbon source for the first pathway is CH$_4$, this 
pathway also creates molecules that react with byproducts 
of CO destruction. Since CO abundance is our focus in this work, we direct our
attention to the second pathway. The reaction network that moves carbon
from CO into COM is sketched in Fig. \ref{fig: carbon_full}. The
formation of COM following CO destruction is primarily driven by helium
ionized by cosmic rays and X-rays. Of the resulting He$^{+}$ ions, more
than half go on to dissociate CO (creating C$^{+}$ and O), while other
He$^{+}$ ions end up ionizing molecules such as H$_2$, C$_2$H$_2$, N$_2$
and CH$_4$. C$^+$ rapidly reacts with CH$_4$ and CH$_3$ to form
hydrocarbon ions. Hydrocarbon ions can go through a series of
charge-exchange reactions until they eventually recombine with an
electron to form neutral hydrocarbons. If the binding energy of the
resulting neutral molecule is large enough, the neutral will quickly
freeze onto the grain surface and remove carbon from the gas-phase
chemistry. Key reactions that initiate hydrocarbon formation are:

\begin{align}
{\rm He}^{+} + \mathrm{CO} &\to \mathrm{O}+\mathrm{C}^{+}+{\rm He}  \label{eq:freeo} \\
\mathrm{C}^{+}+{\rm CH}_{4} &\to \mathrm{C}_{2}\mathrm{H}_{3}^{+} + \mathrm{H} \\
\mathrm{C}^{+}+{\rm CH}_{4} &\to \mathrm{C}_{2}\mathrm{H}_{2}^{+} + \mathrm{H}_2 \\
\mathrm{C}^{+}+{\rm CH}_{3} &\to \mathrm{C}_{2}\mathrm{H}^{+} +\mathrm{H}_{2}.
\label{eq:hcionreactions}
\end{align}
This full reaction chain terminates in frozen-out methanol,
acetaldehyde, and ketene sinks (see Fig. \ref{fig: carbon_full}).

The electrons with which the C$_2$-based hydrocarbon ions eventually
recombine come mostly from cosmic-ray and X-ray ionization of H$_2$.
However, unlike He$^+$, H$_{2}^{+}$ does not contribute to hydrocarbon
formation directly. The majority of H$_{2}^{+}$ initiates HCO$^{+}$
formation through:
\begin{align}
{\rm H}_{2}^{+}+{\rm H}_{2} &\to \mathrm{H}_{3}^{+}+ \mathrm{H} \\
{\rm H}_{3}^{+}+{\rm CO} &\to \mathrm{HCO}^{+} + \mathrm{H}_2. 
\label{eq:h21reactions}
\end{align}
Reactions involving HCO$^+$ often change the charge and/or saturation
of a hydrocarbon (e.g. C$_2$H$_2 +$ HCO$^+ \to {\rm CO} +
\mathrm{C}_2\mathrm{H}_{3}^{+}$), but do not contribute to the initial
formation of the carbon-carbon bond.

After hydrocarbon ions recombine with electrons, the resulting neutral
molecules may freeze onto grain surfaces. The dissociative
recombination reactions
\begin{align}
 \mathrm{C}_{2}\mathrm{H}_{3}^{+} + {\rm e}^{-} &\to \mathrm{C}_{2}\mathrm{H}+2\mathrm{H} \\
 \mathrm{C}_{2}\mathrm{H}_{3}^{+} + {\rm e}^{-} &\to \mathrm{C}_{2}\mathrm{H}_{2}+ \mathrm{H} \\
 \mathrm{C}_{2}\mathrm{H}_{4}^{+} + {\rm e}^{-} &\to \mathrm{C}_{2}\mathrm{H}_{2}+ 2\mathrm{H} \\
 \mathrm{C}_{2}\mathrm{H}_{5}^{+} + {\rm e}^{-} &\to \mathrm{C}_{2}\mathrm{H}_{2}+\mathrm{H}_2 + \mathrm{H} \\
 \mathrm{C}_{2}\mathrm{H}_{5}^{+} + {\rm e}^{-} &\to \mathrm{C}_{2}\mathrm{H}_{3}+ 2\mathrm{H} 
\label{eq:c2hxRecombination}
\end{align}
contribute the most to the total neutral C$_2$H$_x$ abundance, and to
the C$_2$H$_x$ ice budget (recombination pathways are marked with e$^-$
in figure \ref{fig: carbon_full}). As a result, the reaction of CO with
He$^{+}$ starts a chain that moves carbon atoms from CO to hydrocarbons
on million-year timescales.

\subsection{Chemical pathways in the formation of complex organic molecules} 
\label{sec:COM_uncertainty} 

We have seen how the CO abundance beyond 20~AU decreases over time due
to complex organic molecule (COM) formation. Here we trace the
chemical pathways that transfer carbon from CO to COM and assess
whether the outer-disk CO depletion is robust given the construction
of our reaction network.

\subsubsection{Ketene as a carbon sink}
\label{sec:ketene}

Our chemical model does not include any molecule with more than two
carbon atoms. Without more complex species available for reaction
outcomes, H$_2$CCO (ketene) is the most abundant two-carbon molecule
found in our model and serves as a ``sink'' for frozen-out carbon on
the grain surfaces (as seen in Fig. \ref{fig: Chem_combined_1}). The
low binding energy of C$_2$H$_3$ also contributes to the H$_2$CCO
abundance by shutting off the hydrogenation pathway from
C$_{2}$H$_{3}$ to C$_{2}$H$_{6}$ on the grain surfaces and ensuring
that saturation must take place in the gas phase. (The binding
energies of abundant COMs are shown in Table
~\ref{table:IceLine_BindEng}.) Finally, the activation barrier of
$E/k_B = 1210$~K for the reaction $\mathrm{C}_2\mathrm{H}_2 +
\mathrm{H} \to \mathrm{C}_2\mathrm{H}_3$ (Hasegawa et al., 1992),
which has to break the strong C--C triple bond, means that at the cold
temperatures required for acetylene to freeze onto grain surfaces,
hydrogenation proceeds slowly.

The most common reaction for C$_2$H$_3$ in the gas phase is
\begin{equation}
\mathrm{C}_2 \mathrm{H}_3 + \mathrm{H} \to \mathrm{C}_2 \mathrm{H}_2 + \mathrm{H}_2,
\label{eq:c2h3}
\end{equation} 
followed by C$_2$H$_2$ re-freezing onto grain surfaces. A small
fraction of C$_2$H$_3$ gas reacts with ions such as C$_2$H$_{3}^{+}$, HCO$^{+}$, 
and forms more saturated hydrocarbons, then refreezes onto dust
grains:

\begin{align}
\mathrm{C}_2 \mathrm{H}_3 +  \mathrm{C}_2 \mathrm{H}_{3}^{+}  \to \mathrm{C}_2 \mathrm{H}_{5}^{+}+  \mathrm{C}_2 \mathrm{H},\\
\mathrm{C}_2 \mathrm{H}_3 + \mathrm{HCO}^{+}  \to \mathrm{C}_2 \mathrm{H}_{4}^{+} + \mathrm{CO}.
\label{eq:c2h3_others}
\end{align}

At the same time, a small fraction of C$_2$H$_3$ forms a double
bond with free oxygen released in Eq.~\ref {eq:freeo}:
\begin{equation}
\mathrm{C}_2 \mathrm{H}_3 + \mathrm{O} \to \mathrm{H}_2 \mathrm{CCO} + \mathrm{H}.
\label{eq:h2cco}
\end{equation} 
Since H$_2$CCO has very high binding energy (only slightly smaller
than CO$_2$), and we are not including destruction of H$_2$CCO ice
other than desorption, H$_2$CCO ice serves as a sink for COM in our
model. Although the rate for reaction \ref{eq:h2cco} is low, ketene
can nevertheless accumulate on million-year timescales. The reaction
pathway leading to ketene formation is sketched in the lower left part
of Fig. \ref{fig: carbon_full}, with the sink molecule H$_2$CCO
enclosed in a solid box. One should note that ketene is a
representative of the presence of complex organic molecules in our
model. In a real disk, other forms of COM will likely exist.

Densities on grain surfaces are usually much higher than in the gas
phase, so reactions such as hydrogenation on grain surfaces typically
happen very quickly. However, due to the rapid thermal desorption of
C$_2$H$_3$ from the grain surface, more saturated hydrocarbons cannot
be formed efficiently. In most of the disk within $40\AU$, only a
small amount of more saturated C$_2$ hydrocarbons can be formed in the
gas phase through ion-neutral reactions, with a negligible amount
formed though hydrogenation on the grain surface.

\subsubsection{Acetaldehyde as a carbon sink}
\label{acmeth}
Ketene is no longer the sink in regions where the temperature is high
enough for it to be in the gas phase, or where the temperature is low
enough for C$_{2}$H$_{3}$ to stay on the grain surface and continue
hydrogenating to form more saturated hydrocarbons. In those regions,
ice molecules such as CH$_3$CHO (acetaldehyde), CH$_3$OH (methanol),
and C$_2$H$_5$ (ethyl) serve as the carbon sinks, keeping carbon from
re-entering the gas phase.
 
Green lines in Fig. \ref{fig: carbon_full} show the ketene recycling
pathway in a small region below the disk surface between $10-30\AU$,
where the temperature is high enough that H$_2$CCO stays in the gas
phase and the ionization rate is $\ga 10^{-15}$~s$^{-1}$. H$_2$CCO gas
reacts with C$^{+}$ to form CH$_2$CO$^{+}$:
\begin{equation}
\mathrm{C}^{+} + \mathrm{H}_{2}\mathrm{CCO} \to
\mathrm{CH}_{2}\mathrm{CO}^{+} + \mathrm{C}.   
\label{eq:desh2cco}
\end{equation}
CH$_2$CO$^{+}$ then recombines with an electron and undergoes one of
three dissociative recombination reactions to form C$_2$, C$_2$H$_2$,
or CO, with roughly equal branching ratios (see green lines leading
from CH$_2$CO molecule in lower left corner of Fig. \ref{fig:
  carbon_full}). Moreover, due to the high ionization rate, the rate
of reaction \ref{eq:freeo} can be as much as twice the value on the
disk midplane. The abundance of free hydrogen atoms is also high due
to photodissociation of H$_2$ and H$_2$O.  Efficient C$^{+}$
production and high atomic hydrogen abundance speed the production of
C$_2$H$_2$ and allow more saturated hydrocarbons to form on grain
surfaces, despite the volatility of C$_2$H$_3$. The temperature in the
region is high enough for C$_2$H$_5$ to evaporate into the gas phase
once it is formed.  C$_2$H$_5$ then reacts with atomic oxygen to from
the more stable molecule CH$_3$CHO (acetaldehyde), which refreezes on
to the grain surface due to its high binding energy and serves as
another carbon sink:
\begin{equation}
\mathrm{O}+\mathrm{C}_{2}\mathrm{H}_{5} \to {\rm CH}_{3}{\rm CHO} + \mathrm{H}.
\label{eq:ch3cho}
\end{equation}
To summarize, the acetaldehyde-forming reaction pathway
(Fig. \ref{fig: carbon_full}) differs from the ketene-forming pathway
simply because of the warmer temperature that keeps ketene in the gas
phase, the high C$^+$ abundance from reaction \ref{eq:freeo}, and the
high atomic hydrogen abundance. Note that our reaction network does
not include the grain-surface hydrogenation pathway
$\mathrm{C}_2\mathrm{H}_5 + \mathrm{H} \to \mathrm{C}_2\mathrm{H}_6$
(ethane ice). However, \cite{Dodson-Robinson_2009} found low
grain-surface hydrogenation efficiency, leading to significant amounts
of ethane ice only near the acetylene sublimation temperature of 55~K.

\subsubsection{Ethyl and methanol as carbon sinks}
\label{ethylmeth}
In the outer disk where $r > 50$ AU, and after $1\Myr$ in
  most of the disk, C$_2$H$_3$ is able to stay on the grain
  surface and hydrogenate to C$_2$H$_5$, and C$_2$H$_5$ serves as a
  carbon sink.

Due to the lack of C$_2$H$_3$ in the gas phase, and the low
temperature that allows H$_2$CO to freeze out, the reaction 
\begin{equation}
\mathrm{O}+{\rm CH}_{3} \to \mathrm{H}_{2}{\rm CO} + \mathrm{H}
\label{eq:methanol}
\end{equation}
becomes the dominant reaction with atomic oxygen, rather than reaction
\ref{eq:h2cco}. H$_2$CO then freezes onto the grain surface, reacts
with H atoms on the grain surface, and finally forms CH$_3$OH
(methanol), another carbon sink in our model, as follows (letter G
denotes species that are frozen out on grain surfaces):
\begin{align}
\mathrm{GH} +\mathrm{ GH}_{2}{\rm CO} &\to {\rm GCH}_{2}{\rm OH} \\
\mathrm{GH} + {\rm GCH}_{2}{\rm OH} &\to  {\rm GCH}_{3}{\rm OH}.
\label{methanol}
\end{align}

This reaction path explains the evolution of abundances
  of carbon-bearing molecules demonstrated in the lower panel of
  Figure \ref{fig: time_evolution}.  These reactions start to happen
at $r > 60\AU$ where the temperature is about $35\K$, and at $20\AU$
on the midplane at $2.5\Myr$ when the temperature drops below $44\K$.

In summary, as Fig. \ref{fig: Chem_combined_1} and
Fig. \ref{fig: Chem_combined_2} show, H$_2$CCO dominates in most of
the COM-forming region where the temperature is low enough for
H$_2$CCO to remain on the grain surface, but high enough for C$_2$H$_3$
to evaporate into the gas phase. In the small region above the disk
midplane (r $= 10 - 35\AU$, z $= 1-4 \AU$), where the temperature is
high enough for H$_2$CCO and C$_2$H$_5$ to stay in the gas phase and
atomic hydrogen is abundant, CH$_3$CHO serves as the sink for carbon
chemistry. In the outer region where the temperature is low enough for
C$_2$H$_3$ to stay on the grain surface and hydrogenate to C$_2$H$_5$,
we are seeing CH$_3$OH and C$_2$H$_5$ as the end products of COM
chemistry.

The reaction pathways described above demonstrate how H$_2$CCO,
CH$_3$CHO, C$_2$H$_2$ and CH$_3$OH stand in for complex organic
molecules in our model. In reality, organic molecules may grow more
complex as ketene, acetaldehyde, ethyl, and methanol ice react with
other species in ways not included in our model. Despite our upper
limit of two carbon atoms per molecule, the net movement of carbon
from CO to ices should be a robust result for ionized regions that
contain CH$_4$ gas. In order to demonstrate that our reaction network
does not falsely predict carbon depletion from the gas phase, we
investigate an alternative network with a low ketene formation rate in
the next section. We also compare our model abundances with results
from the c2d (cores to disks) {\it Spitzer} Legacy Program as
summarized by \citet{Oberg_2011}, and discuss the
  effect of different assumptions about grain evolution.

\subsection{Model dependence on reaction rates, initial conditions, and grain evolution models}
\label{sec:chem_dependence} 

Our model results depend on the disk physical conditions, initial
cloud phase abundances, and reaction rates.  In this section, we
present a detailed study of how results depend on reaction rates,
initial abundances, and grain evolution.  In section
\ref{sec:lowketene}, we verify that our ketene ice sink does not
artificially remove carbon from the gas phase by running a model with
the ketene formation rate decreased by 10 orders of magnitude. In
Section \ref{sec:c2d}, we compare the abundances from our molecular
cloud preprocessing model with the ice abundances observed by the {\it
  Spitzer} c2d team in the envelopes of low-mass protostars
\citep{Oberg_2011}. In Section \ref{sec:graingrowth},
  we investigate the effects of grain evolution on the disk temperature
  and chemistry.
  
 Our current model cannot consider gaps or inner holes without 
 significant modifications. The effects of such structures on 
chemical evolution requires further study.

\subsubsection{Models with low H$_2$CCO formation rates}
\label{sec:lowketene}

The formation of carbon sinks such as ketene greatly reduces the
gas-phase CO abundance in our model and suggests that observations of
CO depletion may trace complex-molecule formation instead of freezeout
of CO. In this section, we test the robustness of our result by
artificially suppressing ketene formation rates. The CO abundance is
independent of the exact form of carbon-bearing ice as long as our
model does not artificially remove carbon from the gas phase. However,
having H$_2$CCO ice instead of C$_2$H$_x$ ice as a sink in the model
can potentially reduce the amount of oxygen in the gas phase,
therefore affecting the CO abundance. To test the effect of H$_2$CCO
formation rate on CO abundance, we ran the chemical model with
H$_2$CCO formation rates artificially turned down by ten orders of
magnitude while keeping other parameters the same.

We compare output abundances at the end of the $3$ Myr evolution of
this experiment with those from our fiducial COM-forming model
(described in Section \ref{sec: CO2COM}) in Table
~\ref{table:comp_output}. 

The H$_2$CCO formation is strongly suppressed as expected, resulting
in a negligible abundance. Without the pathways that convert
hydrocarbons (C$_2$H$_x$) to ketene, the abundance of hydrocarbons
(C$_2$H$_x$) is significantly higher, 
but most other abundances
show little change. The gas-phase CO abundance increases from 13.6\%
to 17.3\% of the elemental carbon.

To summarize, the choice of end-member species in the chemical
reaction network can affect the gaseous CO abundance. Translating
observed CO isotopologue line intensities into disk surface densities
would include an uncertainty of at least a factor of two for any given
disk annulus. However, the effect of the ketene sink on the CO gas
abundance in our modeled 70~AU disk as a whole is small. One
would be able to translate a measurement of the mass of C$^{17}$O or
C$^{18}$O gas in the planet-forming region of a disk into a total mass
available for giant planet formation without large uncertainties due
to the chemistry of complex organic molecule formation.

\subsubsection{Comparing with the c2d cloud abundances}
\label{sec:c2d} 

We test the sensitivity to initial conditions in
 our chemical reaction network by comparing the
abundances at the end of our molecular cloud preprocessing model with
measured ice abundances in low-mass protostellar envelopes.
\cite{Oberg_2011} combined $\emph{Spitzer}$/ IRS spectra for about 50
low-mass protostars with infrared ice features with near-infrared
ground-based observations of ice features such as $3\um$ H$_2$O,
$4.65\um$ CO, and $3.53\um$ CH$_3$OH, presenting an overview of the
ice inventory during the embedded phase of star formation.  Because
the absolute ice abundances vary from source to source,
\cite{Oberg_2011} presented the median ice abundance ratio with
respect to water ice abundance in the combined sample. The chemical
compositions observed in low-mass protostellar envelopes are expected
to be similar to those at the end of the molecular cloud phase,
because the materials are not yet heavily heated by the central star
or processed by shocks.

We compare c$2$d observed ice abundance ratios and outputs of our
cloud phase model (as described in Section $3.1$) in the first two
columns of Table ~\ref{table:MC_c2d}. The CO$_2$, CH$_4$, and CO
abundances with respect to the water ice abundance in our cloud model
are about 1.2, 5, and 0.5 times the values observed in nearby
star-forming regions, respectively \citep{Oberg_2011}. While not an
exact match, our computed abundances relative to water ice are of the
same order of magnitude as observed values for major carbon-bearing
species.

To test how our disk model varies based on small changes in input
abundances, we conduct an experiment using the \citet{Oberg_2011}
observed abundances as the initial conditions for the chemical
evolution, while keeping other parameters the same. We keep the same
abundances as predicted by our molecular cloud model for H$_{2}$O ice
and minor species not observed by the c$2$d team, but scale the
abundances of other c2d-observed ices to match the c$2$d results. The
abundances predicted by the molecular cloud model, and the abundances
adjusted according to c$2$d ratios, are listed in the last two columns
of Table ~\ref{table:MC_c2d}. This experiment is only to demonstrate
the model dependence on initial input abundances. We do not conserve
the total number of C and O atoms per hydrogen atom by artificially
scaling the molecular abundances.

The change of input abundances does not change CO abundance
significantly in our experiment.  Integrated over the entire disk,
the total available carbon  in the form
  of CO gas increases from 13.6\% to 18.3\%, while the
 carbon locked in COM decreases from 44.8\% to 37.0\% However, the dominant kind of COM
 is different from that of the
standard model. Due to the significantly lower input methane
abundance, the percentage of molecules that contain two carbon atoms is only
one third of the value in the fiducial model, and the methanol
abundance is three times the original value. The paucity of ices that
contain two carbon atoms when the initial methane abundance is low
suggests that even though the formation of the carbon-carbon bond
relies on both CH$_4$ and C$^{+}$ liberated from CO, the formation of
carbon chain depends largely on the methane abundance the disk
inherits from the cloud.

We find that the abundances predicted by our molecular cloud model
agree with observational results within a factor of two. In our
experiment, adjusting the abundances inherited by the disk to match
observed abundances in cold protostar envelopes can lead to a higher
abundance of methanol and a smaller overall COM abundance than in our
fiducial model. However, the computed CO abundance is only 1.35 times higher.

\subsubsection{Different models of  grain evolution}
\label{sec:graingrowth} 
In the fiducial model, we assume $90\%$ by mass of the of dust has already
  grown to larger than mm size, and use a gas to  dust ratio of
  $1000$ throughout. However, grain evolution could be slower than
  what we assume.
 Specifically, the disk could have a larger
  portion of  dust than what we assumed at the beginning
  of the evolution. Small dust grains provide most of the opacity that
  is important for heating and attenuating UV and x-ray radiation, and
  they also provide most of the surface area available for freeze-out
  and grain surface reactions. \cite{Miotello_2014} have found that
  the evolution of dust grains can change the disk opacity, which can
  further increase the significance of isotopologue-selective
  photodissociation, and change the CO fractionation.

To investigate the effect of grain evolution, we construct
  a series of models with gas to  dust ratio gradually evolving
  from 100 to 1000, assuming the same opacity profile for the
  dust grains. We choose to change the gas to dust ratio to trace the 
  loss of solid material to bigger bodies, rather than changing the grain size
  distribution, because a balance between collisional aggregation and 
  fragmentation yields a size distribution whose slope at the small end 
  is insensitive to the dust abundance \citep{Birnstiel_graingrowth_2011}.

In the models considering grain evolution, the total dust
  opacities are larger than those in the fiducial model, resulting in lower disk
  temperature. On average, the disk temperature is $10\%$ lower at $100$ yr
  where the gas-to-dust ratio is $10$ times the fiducial value, and $5\%$ lower at $1\Myr$ when the gas-to-dust ratio is one third of
  the fiducial value.

The output abundances averaged over the entire disk are
  presented in the last column of Table~\ref{table:MC_c2d}. Despite the
  decrease of temperature, the carbon content in the grain evolution
  model is very similar to that in our fiducial model by the end of
  the $3\Myr$ evolution. This is because CO depletion happens on a
  $\Myr$ time scale when grain evolution has already started, and the
  grain surface area is not limiting the freeze-out time scale as long
  as some dust grains are present. The final CO abundance relative to total  
  carbon decreases 
  from 13.6\% in the standard model to 11.8\% in this model, a change 
  in the opposite direction to that found in the other model variations. 
The difference between the two models is largest at 1 Myr, when the CO/C
ratio is 27.5\% in the standard model and 21.8\% in the grain evolution model.
  Unlike in \cite{Miotello_2014}, considering grain evolution does not 
  significantly change abundance ratios of CO isotopologues in our models. 
  We defer detailed discussions on isotopic ratios to section \ref{sec:CO_frac}.

\subsection{Fractionation of CO isotopologues}
\label{sec:CO_frac} 

\begin{figure*}[ht]
\centering
\begin{tabular}{@{}cc@{}}
  \includegraphics[width=0.48\textwidth]{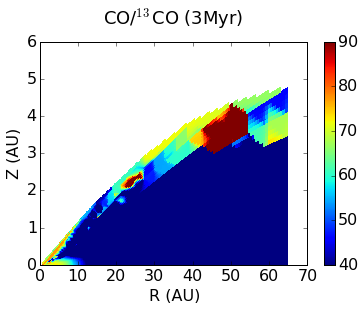} &
  \includegraphics[width=0.48\textwidth]{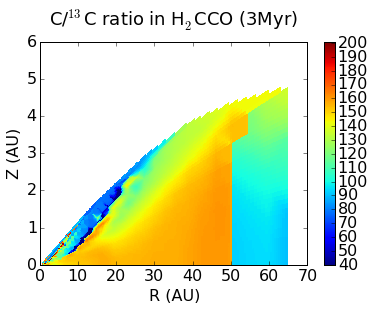} \\
  \includegraphics[width=0.48\textwidth]{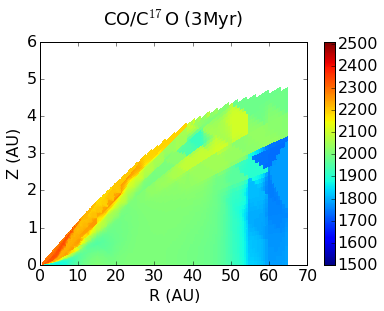} &
    \includegraphics[width=0.48\textwidth]{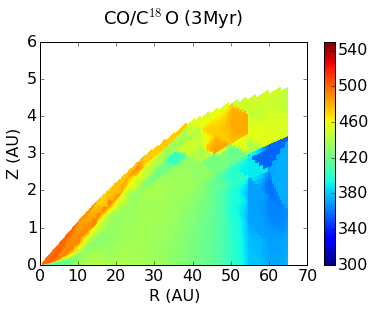} \\
  \end{tabular}
  \caption{ 
  Upper left: CO/$^{13}$CO ratio (all four plots are for the end of the $3\Myr$ evolution);   
  Upper right: C/$^{13}$C ratio in H$_2$CCO ice;
  Lower left: CO/C$^{17}$O ratio;
  Lower right: CO/C$^{18}$O ratio. Unsuperscripted species denote the most common isotopologue.}
    \label{fig: isotopologues}
\end{figure*}

In addition to understanding the formation and destruction pathways of
CO, one also needs to know how C and O isotopes fractionate among
different molecules in order to link $^{13}$CO, C$^{17}$O, and
C$^{18}$O observations with disk masses. The
  isotopologue ratio in CO and the major carbon sink H$_2$CCO are
  presented in Figure \ref{fig: isotopologues}. The CO/$^{13}$CO
ratio is close to the elemental isotopic value of 77.15 in the inner
$15\AU$ where CO gas is abundant. In regions where CO is depleted, the
CO/$^{13}$CO ratio is about $45$, much smaller than the input atomic
abundance ratio. The low CO/$^{13}$CO ratio is inherited from the
molecular cloud model.

In the inner disk, where new CO molecules are forming from recycled
ketene that originally gained its carbon atoms from both CH$_4$ and
CO, the CO/$^{13}$CO ratio approaches the input, atomic C/$^{13}$C
ratio. Integrated over the entire disk, the
 average  CO/$^{13}$CO ratio is  $59.15$. Disk mass estimates
based on optically thin $^{13}$CO rotational emission (possible beyond
$\sim 18$~AU; see Section \ref{sec:opdepth}) should account for
isotopic fractionation.

In the shielded interior of our disk, we predict C$^{16}$O/C$^{17}$O
and C$^{16}$O/C$^{18}$O ratios similar to those measured by
\cite{smith09} from observations of the envelope of protostar
Reipurth~50. Our computed CO/C$^{17}$O and CO/C$^{18}$O ratios are
slightly smaller than the input $^{16}\mathrm{O}/^{17}\mathrm{O} =
2300$ and $^{16}\mathrm{O}/^{18}\mathrm{O} = 500$ ratios throughout
most of the disk, though the recycling of ketene back to CO pushes
C$^{16}$O/C$^{17}$O and C$^{16}$O/C$^{18}$O higher near the disk
surface. The CO oxygen isotope fractionation identified in our model
has an average effect of less than $14\%$ (averaged
over mass). We find that oxygen isotope fractionation would not
significantly bias disk mass estimates based on C$^{17}$O or C$^{18}$O
emission. 

We also caution against over-interpretation of the oxygen
isotope fractionation found in this model, as variations in UV flux
that are well within observed ranges \citep{France_UV_2014} can affect
isotope ratios. In \cite{Miotello_2014}, UV radiation is not attenuated efficiently in
  disks with only larger grains. Rare CO isotopologues only exist in a
  smaller region further below the disk surface due to photo-dissociation, 
  resulting in larger CO to rare CO ratios. In contrast, our models 
  include a population of small dust grains even for models with grain evolution 
  (Section~\ref{sec:graingrowth}), and therefore the UV radiation is sufficiently
  absorbed. Moreover, following \cite{Landry_2013}, our disk models only
  follow the disk up to about two scale heights, and do not follow the
  disk atmosphere, where photochemical reactions matter the most. This
  explains why unlike in \cite{Miotello_2014}, considering grain
  evolution does not significantly change the CO isotopologue ratios.

Having verified that our predicted $^{13}$CO, C$^{18}$O, and C$^{17}$O
abundances are robust results that (1) do not depend heavily on our
reaction network choice and (2) closely track observed values, we are
now ready to calculate the optical depths of rotational emission
lines.

\section{Optical depth of rotational emission from CO isotopologues}
\label{sec:opdepth}

Our goal is to demonstrate that rare CO isotopologues can provide a
window into the planet formation zone through their rotational
emission lines. The ideal emission line would be optically thin all
the way to the disk midplane, but still detectable with ALMA.  We
calculate the optical depth of CO rotational emission lines at each
disk radius by integrating the absorption efficiency through the disk.
We first calculate the total absorption efficiency across the emission
line as
\begin{equation}
\int k_{\nu}d\nu = \frac{h\nu_{ij}}{c}\left(n_iB_{ij}-n_jB_{ji}\right),
\label{eq:abseff}
\end{equation}
in which
\begin{align}
B_{ji} &= \frac{g_i}{g_j}B_{ij}, \\
B_{ij} &= \frac{8\pi^3}{3h^2}|\mu_{ij}|^2, \\
n_i &= f_in_x.
\label{eq:einsteinB}
\end{align}
$k_{\nu}$ is the absorption coefficient, $n_i$, and $n_j$ are the
number of molecules in states i and j (i is the lower state), and
$f_i$ and $f_j$ are the fraction of molecules in states i and j. $g_i$
and $g_j$ are the degeneracies; $\nu_{ij}$ is the frequency of the
transition; $\mu_{ij}$ is the matrix element of the electric dipole
moment for the transition; and $B_{ij}$ and $B_{ji}$ are the Einstein
coefficients of the transition.  Substituting into equation
\ref{eq:abseff}, we have:
\begin{equation}
\int k_{\nu}d\nu = \frac{h\nu_{ij}}{c} f_in_x \frac{8\pi^3}{3h^2}|\mu_{ij}|^2\left[1-exp\left(\frac{-h\nu_{ij}}{kT}\right)\right].
\label{eq:totabseff}
\end{equation}
We take $f_i$ as the local thermodynamic equilibrium (LTE) value set by the Boltzmann distribution.

To estimate the absorption at the line center $k_0$, we divide the
total absorption efficiency by the full-width half-maximum $\Delta\nu$
of the Doppler broadening profile:
\begin{equation}
\Delta\nu=2.355\times\sqrt{\frac{\nu_0^2kT}{Mc^2}},
\label{eq:dopplerbroadening}
\end{equation}
where $M$ is the mass of the molecule. We then have
\begin{equation}
k_{0}= \frac{\int k_{\nu}d\nu}{\Delta\nu}.
\label{eq:linecenteropacity}
\end{equation}
Finally, we compute face-on optical depths as a function of distance from the star by integrating the
absorption efficiency $k_0$ over the disk's vertical direction as
\begin{equation}
\tau_{0}=\int_{0}^{z_{surface}}k_{0}dz.
\label{eq:opdepth}
\end{equation}

\begin{figure*}[ht]
\centering
\begin{tabular}{@{}cc@{}}
  \includegraphics[width=0.40\textwidth, angle = 90]{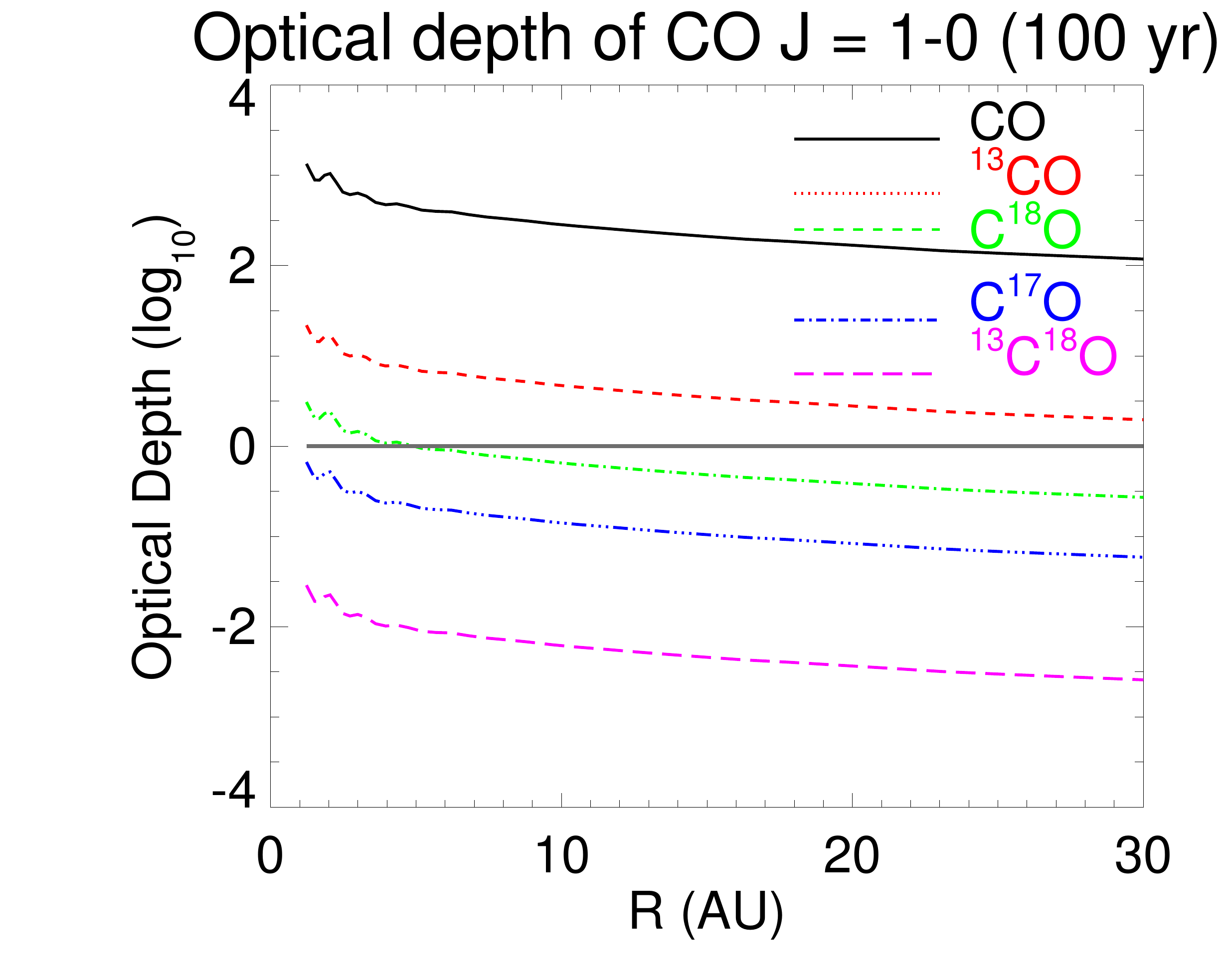} &
  \includegraphics[width=0.40\textwidth, angle = 90]{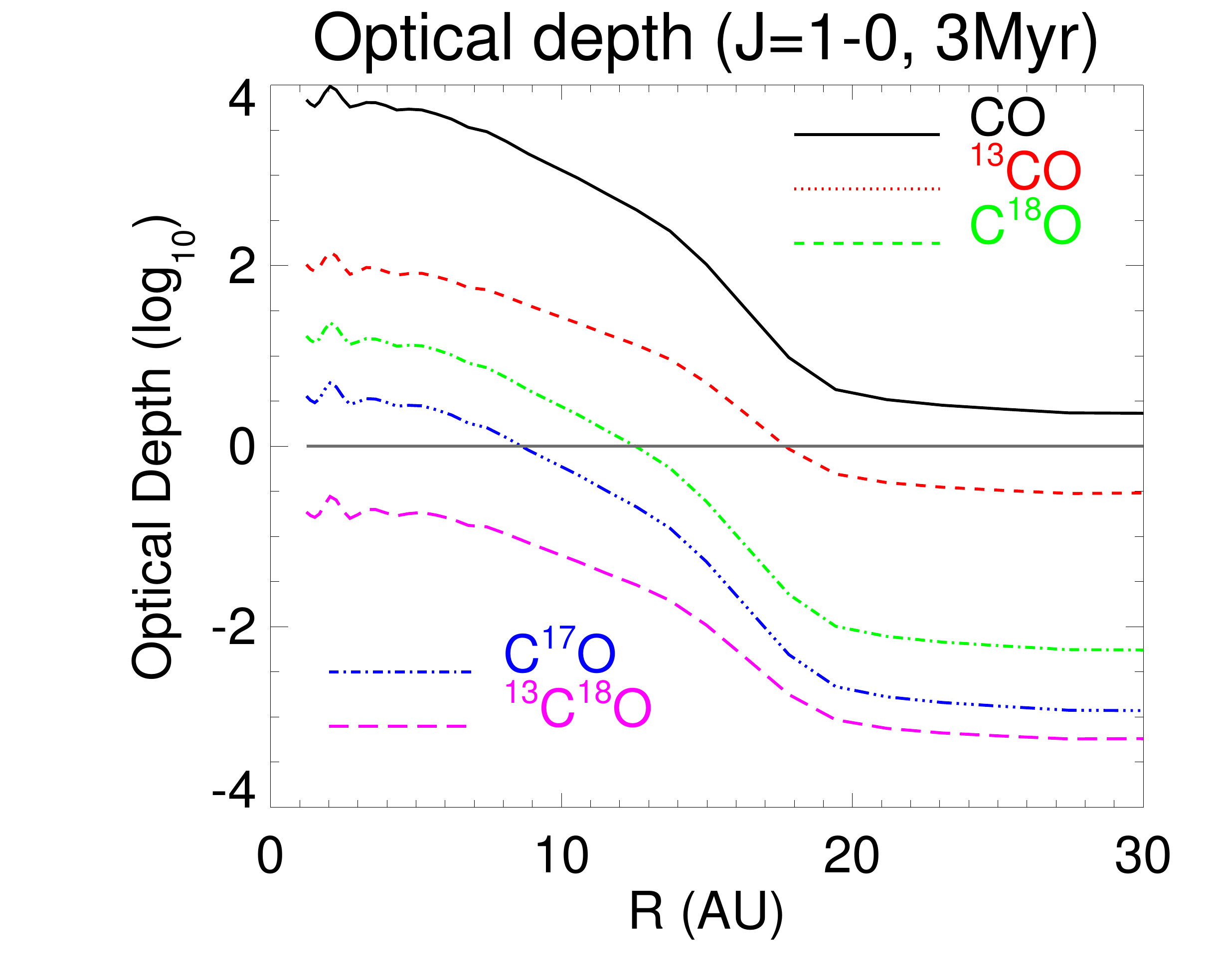} \\
  \includegraphics[width=0.40\textwidth, angle =90]{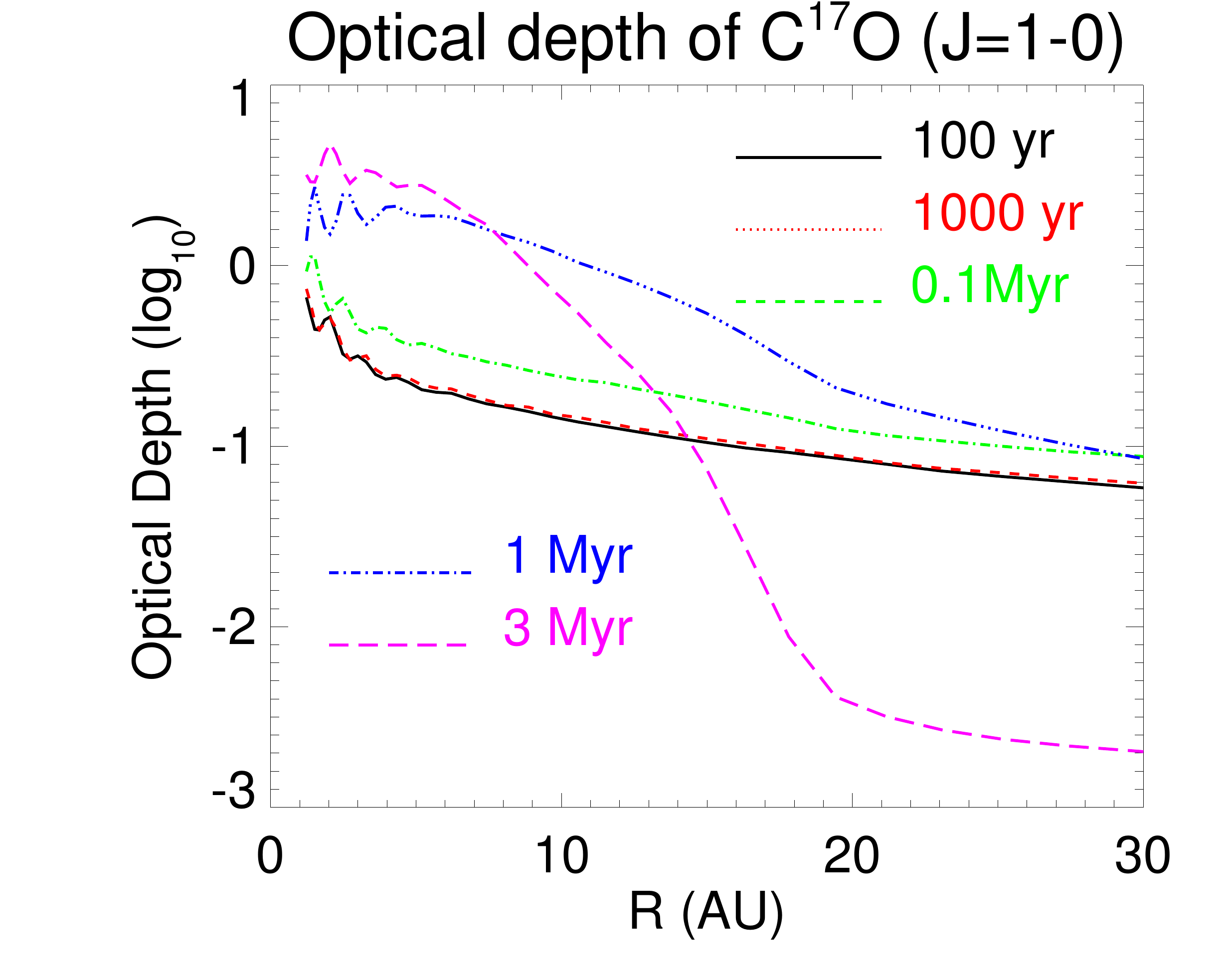} &
  \includegraphics[width=0.50\textwidth]{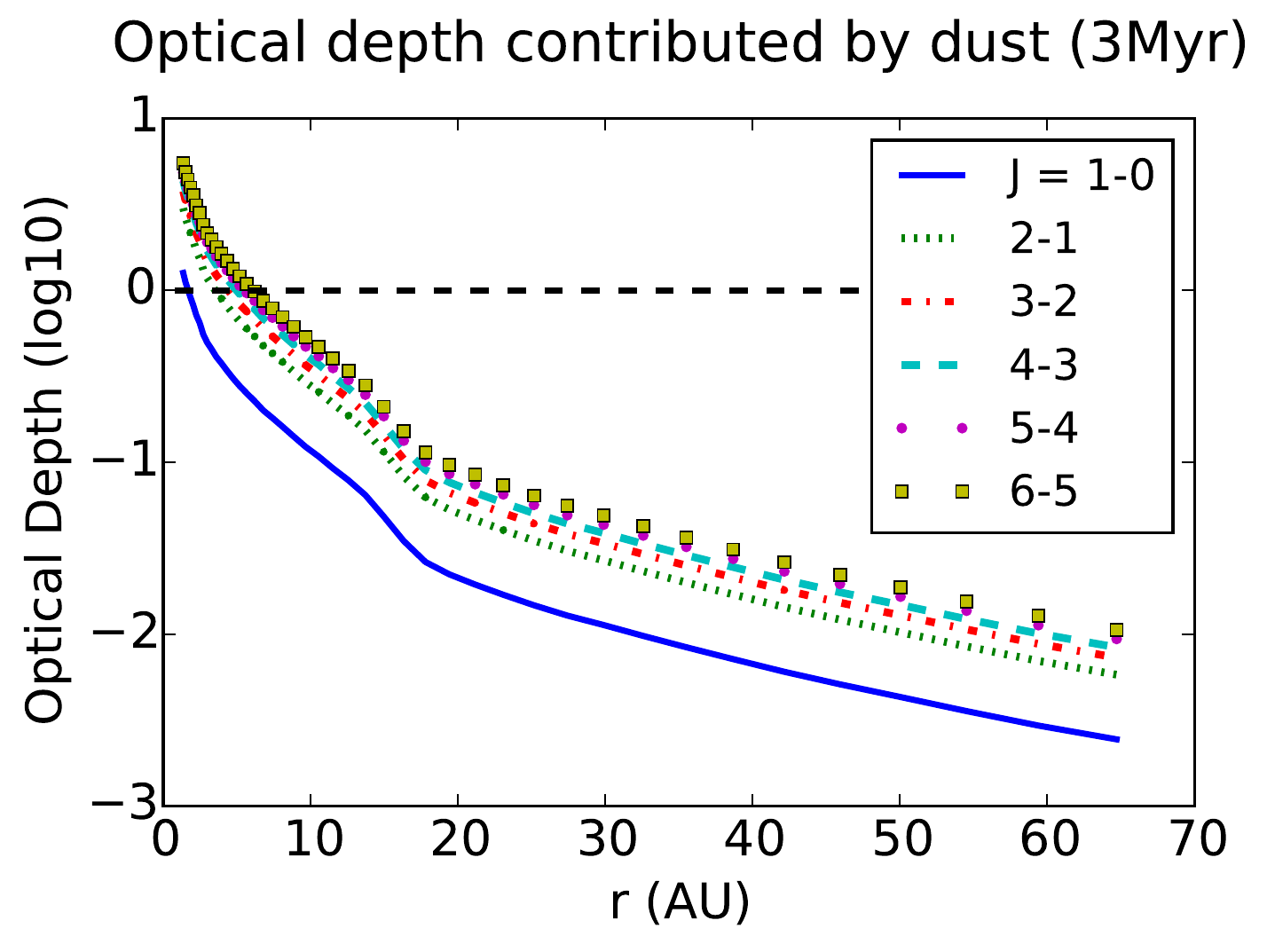} \\
  \end{tabular}
  \caption{Upper left panel: Optical depth of various CO isotopologues
    J $= 1\to 0$ ($100$ yr).  Upper right panel: Optical depth of
    various CO isotopologues J $= 1\to 0$ ($3$ Myr).  Lower left
    panel: Optical depth of C$^{17}$O J $= 1\to 0$ (time
    evolution). The optical depth increases over time roughly in r $<
    20$ AU due to the photodissociation of CO$_{2}$ lead by cosmic
    ray-induced photons, and the optical depth at r $> 20$ AU
    decreases over time due to the formation of COMs.  Lower right
    panel: Optical depth contributed by dust emission at the
    wavelength of various CO transitions. Dust emission should not
    contribute much to observed fluxes in low-order CO rotational
    emission lines beyond $10\AU$.}
    \label{fig: Optical_depth_combined}
\end{figure*}

The optical depths of J$=1 \to 0$ lines for CO isotopologues are shown
in the upper panels of Figure \ref{fig: Optical_depth_combined}. The
upper-left panel shows optical depths after 100~years of disk
evolution (i.e., a disk surrounding a just-emerged T-Tauri star), and
the upper-right panel shows the disk after 3~Myr of evolution.  We
highlight the J$=1 \to 0$ transition because the temperature of CO gas
is relatively high compared to the energy needed to excite the J $=1$
rotational level ($5.5K$), therefore J $= 1\to 0$ line is the most
optically thin line among low order rotational lines observable with
ALMA. We can immediately see that C$^{17}$O traces the
  disk midplane outside $8$~AU, and C$^{18}$O traces the midplane
  outside $12$~AU. Observations of C$^{17}$O and C$^{18}$O emission
from nearby disks would provide estimates of the amount of mass
available to form planets like Uranus and Neptune, and place a lower
limit on the mass available to form Jupiter and Saturn.

The optical depth of C$^{17}$O as a function of radius is plotted for
several epochs in the lower left panel of Figure \ref{fig:
  Optical_depth_combined}. The optical depth is
a strong function of radius and time.
The optical depth in the inner part of the disk
  increases as the disk evolves due to the dissociation of CO$_2$ into
  CO, while the optical depth beyond $15\AU$ first increases,
but then decreases and falls
below the initial value due to the formation of more complex
carbon-bearing molecules (sections \ref{sec:CO}, \ref{sec:
  CO2COM}). At $3\Myr$ the optical depth remains above
  one out to about $42\AU$ for CO, $18\AU$ for $^{13}$CO, $12\AU$ for
  C$^{18}$O, and $8\AU$ for C$^{17}$O. The possibility that CO
abundance changes over time due to chemical reactions means that disk
masses measured by CO emission are degenerate with age.

Continuum emission from dust also contributes flux at the frequencies
of CO emission lines. Even if the emission from a CO isotopologue is
optically thin, high optical depth in the dust could prevent us from
seeing the disk midplane. To estimate the optical depth contributed by
the dust, we integrate the dust opacity at the relevant wavelength
over the disk height. We use opacities from
\cite{Semenov_Henning_2003}, for grains with a 5-layered sphere
topology. As in the thermodynamic model (see Section
\ref{sec:thermodynamic} for details), we use a gas to dust mass
ratio of $1000$. We assume the dust has undergone some degree of grain
growth, with remaining solids locked into larger bodies, consistent
with our RADMC dust radiative transfer models. Optical depth
contributed by dust is plotted in the lower right panel of Figure
\ref{fig: Optical_depth_combined}. The optical depths contributed by
dust are almost the same for all isotopologues due to the small shift
in frequency between emission lines, so we are only showing the
frequency at the center of CO transitions. The optical depths
contributed by dust are smallest for the J$=1\to0$ line due to lower
frequency - less than one beyond $2\AU$, and less than $0.1$ beyond
$10\AU$. The dust optical depth for higher order transitions become
less then one within $10\AU$.

For low-mass disks that have experienced some grain growth, C$^{17}$O
is a promising tracer of the disk midplane in the giant planet-forming
region. Outside of $10$ AU, where analogs to the Kuiper Belt may be
forming \citep{Bryden_DebrisDisk_2009}, C$^{18}$O and $^{13}$CO may be
useful midplane tracers. Note, however, that our fiducial model is a
low-mass disk ($0.015 \Msun$), and the actual midplane locations
traced by CO gas may therefore fall somewhat outside of $10$
AU. However, our results demonstrate the value of observing CO
isotopologues to reconstruct the mass distributions in the inner $30$
AU of nearby disks.

\section{Conclusions}

Our chemical model of an MRI-active
protoplanetary disk has led to the following conclusions.

CO does not freeze out anywhere in our modeled region - the inner
$70\AU$ of the disk - due to efficient heating by stellar
irradiation. Instead, the abundance of CO is a complex function
of both radius and time. This dependence must be modeled correctly
in order to deduce gas properties from observations of CO isotopologues.

The fate of CO is tied up in the formation of complex organic
molecules.  While the detailed chemical evolution depends on the input
abundances from the molecular cloud and which species are included in
the network, the main result is robust to these variations. Different models 
of grain evolution also produce small changes in the outcomes. For three 
different chemical assumptions and two different scenarios
for grain evolution, the partition of carbon, integrated over the
modeled inner $70$ AU of the disk, ranges from $11.8\%$ to $18.3$\% in
CO gas, $31.0$\% to $38.4$\% in CO$_2$ ice, and $37.0\%$ to $45.0$\%
in complex organic molecules.

Fractionation of oxygen isotopes appears not to play a major
role in C$^{17}$O abundances, though stars with unusually
strong UV accretion luminosity may have disks depleted in C$^{17}$O. 
The optical depth of low-J rotational lines of C$^{17}$O are around unity
in the giant planet forming region, while more common isotopologues
are quite opaque. With our computed C$^{17}$O/H$_2$ abundance as a
function of radius and an age, one would be
able to translate the observed C$^{17}$O line intensity into available
planet-forming mass, to within a factor of a few.
The optical depth of dust in the CO isotopologue (J$=1 \to 0$) transition
wavelengths is small enough to allow observations of the mid-plane beyond
a few AU; for higher transitions the dust will be opaque within 5-8 AU. 
However, this result relies on our assumption of grain evolution,
in which 90$\%$ of the solid mass has accumulated in pebbles and larger
solid objects at the start of disk evolution at 0.1 $\Myr$, but
our experiment with slower evolution of the gas to dust ratio produce
similar results in the end.

Looking ahead to future work, we emphasize that emission lines from
the CO isotopologues all have different optical depths and can probe
different vertical layers of a target disk. Comparing line profiles of
emission from multiple isotopologues could reveal vertical variations
in turbulent velocity, and therefore constrain the angular momentum
transfer mechanism that drives the disk evolution. The current leading
angular momentum transfer mechanism---MRI turbulence---predicts
non-turbulent ``dead'' zones on the disk midplane, which leads to a
decrease of the turbulent velocity toward the midplane
(\cite{Fromang_Nelson_2006}, \cite{Simon_2011}). However, new models
by \cite{Bai_Wind1_2013} and \cite{Gressel_2015} suggest that
magnetocentrifugal winds are more likely to drive accretion than the
MRI. Aside from possibly driving accretion, turbulence also determines
the behavior of dust grains to a great extent. An observational
investigation of gas velocities in the midplanes of planet-forming
disks would have profound implications for planetesimal growth
models. ALMA is able to detect emission from multiple CO isotopologues
in one observation. With its new long baselines and large collecting
area, it is sensitive to the inner regions of disks, which have small
emitting areas but contribute appreciably to line wings. CO
observations may be able to probe not only the mass distribution in
planet-forming regions, but the gas dynamics as well.

We conclude that CO isotopologue rotational emission can probe
the mass distribution and chemical evolution of the inner radii of
protostellar disks, allowing observers to peer directly into planet
nurseries.

Acknowledgments: Work by MY, KW, SDR and NJT was supported by NASA
grant NNX10AH28G. NJE and MY were supported in part by NSF Grant
AST-1109116 to the University of Texas at Austin. This work was
performed in part at the Jet Propulsion Laboratory, California
Institute of Technology. NJT was supported by grant 13-OSS13-0114 from
the NASA Origins of Solar Systems program. We are grateful to the referee 
for helpful suggestions. We would like to thank
Edwin Bergin, Jacob Simon, Ilse Cleeves, Jeong-Eun Lee, Seok-Ho Lee,
Jeffrey Cuzzi, Paul Estrada, Colette Salyk, Karin \"{O}berg, and
Raquel Salmeron for useful discussions.

\addcontentsline{toc}{chapter}{Bibliography} 
\bibliographystyle{aasjournal}
\bibliography{references}

\clearpage


\renewcommand{\arraystretch}{0.9}

\begin{table*}[!ht]
\small

\caption{A summary of parameters and definitions of variables 
\label{table:VarList}}
\begin{center}
\begin{tabular}{p{2cm} | p{12cm}}

\hline\hline 
Symbol  & Definition or Value\\
\hline
ISRF & Interstellar radiation field \\
&  $1 \rm ISRF = 1.6\times 10^{-3} \rm ergs^{-1}cm^{-2}$ for the ionizing UV radiation \\
COM & Complex organic molecules \\
MRI & Magneto-rotational instability \\
\hline
Stellar mass & $0.95\Msun$, we follow the stellar evolution from $0.1\Myr$ ($\Tstr=4600K$, and $\Rstr=5.5\Rsun$), to $3\Myr$ ($\Tstr=4500K$, and $\Rstr=1.5\Rsun$). \\

Disk mass & $0.015\Msun$ within $70\AU$ from the central star \\

n$_{\rm H_2}$ & number density of hydrogen molecules \\ 
f$_{\rm H}$ & 0.735, mass fraction of hydrogen \\ 
Abundance - A(B) & number density with respect to the number density of hydrogen
nuclei (n$_{\rm H}+2\rm n_{\rm H_{2}}$). A(B) stands for
A$\times{10}^B$ \\ 

\hline

T & disk temperature (same for dust and gas)
\\ T$_{eq}$ & equilibrium temperature contributed by stellar irradiation \\ 
T$_{acc}$ & accretion temperature contributed by the MRI turbulence \\
\hline 

$\zeta_{abs} $ & $6\times{10}^{-12}$~s$^{-1}$, ionization rate for direct x-ray absorption \\
$\zeta_{sca}$ &  ${10}^{-15}$~s$^{-1}$, ionization rate for scattered x-ray photons \\ 
$L_{X,29}$ & $20$, stellar X-ray luminosity in units of $10^{29}\: ergs^{-1}$   \\ 

\hline $\Sigma_{cr}$ & $96$~gcm$^{-2}$, characteristic column density of attenuation for cosmic rays \\ 
$\zeta_{cr} $ & cosmic ray ionization rate, $1.2\times{10}^{-17}$~s$^{-1}$ for H$_{2}$ molecules
when calculating the reactions lead by cosmic ray induced photons, the
total ionization rate is taken as $1.3\times{10}^{-17}$~s$^{-1}$. \\

\hline\hline
\end{tabular}
\end{center}
\small
\vspace{0.3cm}
\end{table*}

\begin{table*}[!ht]
\caption{Input abundances of the molecular cloud model
\label{table:MC_input}}

\begin{center}
\small
\begin{tabular}{|l|c||l|c|}
\hline\hline 
Species & Abundance $^{1, 2}$ & Element &Abundance\\
\hline
 H & 1.00 (-2) & He & 1.40 (-1)\\ 
 C$^+$ & 7.21 (-5) & $^{13}$C$^+$ & 9.34 (-7)\\ 
 O & 1.76 (-4) &  $^{18}$O & 3.51 (-7)\\ 
$^{17}$O & 7.64 (-8) &  N & 2.14 (-5)\\ 
Si$^{+}$ & 3.00 (-9) &  & \\ 
\hline\hline
\end{tabular}
\end{center}
\small
\vspace{0.3cm}
$^1$ Number density with respect to the number density of hydrogen nuclei  (n$_{\rm
  H}+2\rm n_{\rm H_{2}}$). The abundances are taken from \cite{Graedel_1982_MCinput}. \\
$^2$ A(B) stands for A$\times{10}^B$
\vspace{0.3cm}
\end{table*}

\renewcommand{\arraystretch}{1}
\begin{table*}[!ht]
\caption{Binding energies and ice locations for carbon-bearing molecules
\label{table:IceLine_BindEng}}
\begin{center}
\small
\begin{tabular}{|l|c|c|c|}
\hline\hline 

Species &Binding Energy (K) $^{1, 2}$& Ice line ($100$ yr) $^3$	& Ice line ($3\Myr$) $^3$	
\\
\hline 
H$_2$O &	5.77 (3)  & 	$1.5\AU$ &	$1\AU$\\
CO$_2$ & 	2.86 (3) & 	$18\AU$ &       $2\AU$\\
CO 	     &        8.55 (2) & 	 $>70\AU$ &	$>70\AU$\\
CH$_4$ &	1.08 (3) & 	$>70\AU$ &	$>70\AU$\\
\hline
Species &Binding Energy (K) &   \multicolumn{2}{|c|}{Ice location ($3\Myr$) $^4$}\\
\hline
CH$_3$  &        1.16 (3) &   \multicolumn{2}{|c|}{} \\
CH$_3$OH  &   4.23 (3) &   \multicolumn{2}{|c|}{r$=5-20$AU, z=$0.2-2$AU, and r$>40$AU }\\

C$_2$H$_2$  &  2.40 (3) &   \multicolumn{2}{|c|}{ $2-25\AU$  surface}\\
C$_2$H$_3$  & 1.76 (3) &   \multicolumn{2}{|c|}{}\\
C$_2$H$_4$  & 2.01 (3) &   \multicolumn{2}{|c|}{} \\
C$_2$H$_5$  & 2.11 (3) &   \multicolumn{2}{|c|}{$>40$AU}\\
C$_2$H$_6$  &  2.32 (3)&   \multicolumn{2}{|c|}{}\\
H$_2$CCO  & 2.52 (3) &   \multicolumn{2}{|c|}{most of the disk $5-45$AU}\\
CH$_3$CHO  & 2.87 (3) &   \multicolumn{2}{|c|}{r$=8-25$AU, z$=0.5-2.5$AU}\\
H$_2$CO  & 1.76 (3) &   \multicolumn{2}{|c|}{}\\

\hline\hline
\end{tabular}
\end{center}
\small
\vspace{0.3cm} $^1$ Binding energy is cited in energy (E in erg)
divided by Boltzmann's constant (k$_{B}$ in erg/K) for easy
interpretation, therefore has a unit of Kelvin. \\ $^2$ A(B) stands
for A$\times{10}^B$\\ $^3$ The ice line (condensation front) location
on the disk midplane. \\ $^4$ For CH$_3$OH, C$_2$H$_2$, C$_2$H$_5$,
H$_2$CCO, and CH$_3$CHO, we mark the location where we see certain ice
species, which depend on both the condensation temperature and the
ability to form the molecule. The change of abundance can be
  very gradual. Here we quote the boundary location where $10\%$ of
  carbon is stored in corresponding species. We are only showing
results for the end of evolution ($3\Myr$) because those molecules are
not formed until about $1\Myr$ into the disk evolution. \\

\vspace{0.3cm}
\end{table*}

\begin{table*}[!ht]
\caption{Abundances at the end of the cloud phase
\label{table:MC_c2d}}
\begin{center}
\small
\begin{tabular}{l|c|c|c|c}
\hline\hline 
Ice species&  MC fraction $^1$ & c2d fraction  $^2$ & MC abundance $^3$ & c2d abundance $^4$\\
\hline
H$_{2}$O        & 100& 100 & $9.1(-5)$     & $9.1(-5)$ \\
CO$_{2}$         & 35 & 28   & $3.2(-5)$   &    $2.7 (-5)$\\
CH$_4$          &  24 & 5   & $2.2(-5)$   &   $4.6 (-6)$  \\
CO                 & 15 & 29   & $1.3(-5)$  &     $2.7 (-5)$  \\
\hline
Gas species & &\\
\hline
 CO            &  & NA  $^5$       & $1.9 (-6)$  & $1.9 (-6)$ \\
\hline\hline
\end{tabular}
\end{center}
\small
\vspace{0.3cm}

$^1$ Fractional abundances in our molecular cloud model, normalized to
the H$_2$O ice abundance.\\
$^2$ Fractional abundances in the c2d result, normalized to
the H$_2$O ice abundance. \\
$^3$ Output abundances in our molecular cloud model.\\
$^4$ Abundances scaled according to the c2d result, using the H$_2$O
abundance as the reference.\\
$^5$ \cite{Oberg_2011} does not include gaseous species.\\
\vspace{0.3cm}
\end{table*}

\begin{table*}[!ht]
\caption{Output abundances from different models
\label{table:comp_output}}

\begin{center}
\small
\begin{tabular}{l|c|c|c|c}
\hline\hline 
Ice    &  Standard$^1$  &  Low H$_2$CCO$^2$  &c2d$^3$  & evolving D/G \\
\hline

CO$_{2}$         &  $36.2$    & $37.6$  & $31.0$         & $38.4$\\
CH$_3$OH     &  $8.1$     &  $7.4$  & $24.6$   & $10.1$\\
CH$_3$CHO  &    $3.9$     &    $4.4$   & $5.6$ &  $3.01$\\
H$_2$CCO     & $15.2$        &  $0.0$  & $3.4$   &  $9.0$\\
C$_2$H$_x$ $^4$  & $16.1$&   $25.9$  &  $3.1$  &  $21.0$\\
COM $^5$  & $44.8$&    $39.4$  & $37.0$  &$45.0$\\
\hline \hline
Gas  &&&&\\
\hline

CO                 &  $13.6$     &  $17.3$ &  $18.3$  &$11.8$\\ 
CH$_4$          &  $1.6$     &$1.7$ &  $0.2$   & $1.9$\\
\hline\hline
\end{tabular}
\end{center}
\small
\vspace{0.3cm} $^{1}$ Model with original reaction rates and the input
from our cloud model abundances. The percentage of total available
carbon contained in each species is shown in the table. All the
results are values integrated over the entire disk at the end of our
modeled disk evolution at $3\Myr$. \\ 
$^{2}$ Model with small H$_2$CCO formation rates and the input from our cloud model
abundances. \\
$^{3}$ Model with original
reaction rates and the input adjusted to the c2d fractional
abundances. See Table \ref{table:MC_c2d} for details.\\ 
$^4$ Hydrocarbons that contain two carbon
atoms.
\\ $^5$ The total abundance of complex organic molecules,
including CH$_4$, CH$_3$OH, H$_2$CCO, CH$_3$CHO, and C$_2$H$_x$.\\
\vspace{0.3cm}
\end{table*}


\clearpage

\end{document}